\documentclass[prd,superscriptaddress,amsfonts,amssymb,amsmath,showpacs,twocolumn]{revtex4-2}
\usepackage{bm}
\usepackage{amsfonts}
\usepackage{latexsym}
\usepackage[latin1]{inputenc}
\usepackage{graphicx}
\usepackage{amsmath}
\usepackage{palatino}
\usepackage{mathpazo}
\usepackage{textcomp}
\usepackage{float}
\usepackage{booktabs}
\usepackage{dcolumn}
\usepackage{ragged2e}
\usepackage{hyperref}
\usepackage[dvipsnames]{xcolor}
\hypersetup{colorlinks,citecolor=OrangeRed}
\hypersetup{colorlinks=true, linkcolor=blue, urlcolor=Sepia}
\usepackage{amsmath}
\usepackage{xcolor}
\usepackage{orcidlink}
\usepackage[caption=false]{subfig}
\usepackage{commath}
\def\jnl@style{\it}
\def\aaref@jnl#1{{\jnl@style#1}}

\def\aaref@jnl#1{{\jnl@style#1}}

\def\aj{\aaref@jnl{AJ}}                   % Astronomical Journal
\def\apj{\aaref@jnl{ApJ}}                 % Astrophysical Journal
\def\apjl{\aaref@jnl{ApJ}}                % Astrophysical Journal, Letters
\def\apjs{\aaref@jnl{ApJS}}               % Astrophysical Journal, Supplement
\def\apss{\aaref@jnl{Ap\&SS}}             % Astrophysics and Space Science
\def\aap{\aaref@jnl{A\&A}}                % Astronomy and Astrophysics
\def\aapr{\aaref@jnl{A\&A~Rev.}}          % Astronomy and Astrophysics Reviews
\def\aaps{\aaref@jnl{A\&AS}}              % Astronomy and Astrophysics, Supplement
\def\mnras{\aaref@jnl{Mon.~Not.~Roy.~Astron.~Soc.}}             % Monthly Notices of the RAS
\def\prd{\aaref@jnl{Phys.~Rev.~D}}        % Physical Review D
\def\plb{\aaref@jnl{Phys.~Lett.~B}}        % Physics Letters B
\def\prc{\aaref@jnl{Phys.~Rev.~C}}  % Physical Review C
\def\prl{\aaref@jnl{Phys.~Rev.~Lett.}}    % Physical Review Letters
\def\qjras{\aaref@jnl{QJRAS}}             % Quarterly Journal of the RAS
\def\skytel{\aaref@jnl{S\&T}}             % Sky and Telescope
\def\ssr{\aaref@jnl{Space~Sci.~Rev.}}     % Space Science Reviews
\def\zap{\aaref@jnl{ZAp}}                 % Zeitschrift fuer Astrophysik
\def\nat{\aaref@jnl{Nature}}              % Nature
\def\aplett{\aaref@jnl{Astrophys.~Lett.}} % Astrophysics Letters
\def\apspr{\aaref@jnl{Astrophys.~Space~Phys.~Res.}} % Astrophysics Space Physics Research
\def\physrep{\aaref@jnl{Phys.~Rep.}}      % Physics Reports
\def\physscr{\aaref@jnl{Phys.~Scr}}       % Physica Scripta
\def\commat{\aaref@jnl{Comm.~Math.~Phys.}}              % Communications in Mathematical Physics
\def\science{\aaref@jnl{Science}}               % Science
\def\cqg{\aaref@jnl{Classical Quant.~Grav.}}            % Classical and Quantum Gravity
\def\jpcs{\aaref@jnl{JPCS}}                                     % Journal of Physics Conference Series
\def\ijmpd{\aaref@jnl{Int.~J.~Mod.~Phys.~D}}                    % International Journal of Modern Physics D
\def\grg{\aaref@jnl{Gen.~Relat.~Gravit.}}               % General Relativity and Gravitation
\def\rpp{\aaref@jnl{Rep.~Prog.~Phys.}}          % Reports on Progress in Physics
\def\npa{\aaref@jnl{Nucl.~Phys.~A}}        % Nuclear Physics A
\def\lrr{\aaref@jnl{Living Rev.~Rel.}}                   % Living reviews in relativity
\def\jcap{\aaref@jnl{J.~Cosmology Astropart.~Phys.}}    % Journal of cosmology and astroparticle physics
\def\rmp{\aaref@jnl{Rev.~Mod.~Phys.}}   %Reviews of modern physics
\def\epjc{\aaref@jnl{Eur.~Phys.~J.~C}}

\allowdisplaybreaks[1]
\begin{document}
%\color{red}
\color{black}       %% For one column
%
%Anisotropic stars from interacting quark matter equation of state
\title{Effects of anisotropic pressure on interacting quark star structure}

\author{Juan M. Z. Pretel \orcidlink{0000-0003-0883-3851}}
 \email{juanzarate@cbpf.br}
 \affiliation{
 Centro Brasileiro de Pesquisas F{\'i}sicas, Rua Dr.~Xavier Sigaud, 150 URCA, Rio de Janeiro CEP 22290-180, RJ, Brazil
}
\affiliation{
 Grupo de investigaci{\'o}n ``Gravitaci{\'o}n, Cosmolog{\'i}a, Campos y Cuerdas'', 
 Universidad Nacional de Trujillo, Departamento de F{\'i}sica, Av.~Juan Pablo II S/N; Ciudad Universitaria, Trujillo, La Libertad, Peru
}

\author{Takol Tangphati \orcidlink{0000-0002-6818-8404}} 
\email[]{takoltang@gmail.com}
\affiliation{School of Science, Walailak University, Thasala, \\Nakhon Si Thammarat, 80160, Thailand
}

\author{Ayan Banerjee \orcidlink{0000-0003-3422-8233}} 
\email{ayanbanerjeemath@gmail.com}
\affiliation{Astrophysics and Cosmology Research Unit, School of Mathematics, Statistics and Computer Science, University of KwaZulu--Natal, Private Bag X54001, Durban 4000, South Africa
}

\author{Anirudh Pradhan \orcidlink{0000-0002-1932-8431}} \email[]{pradhan.anirudh@gmail.com}
\affiliation{Department of Mathematics, Institute of Applied Sciences and Humanities, GLA University, Mathura-281 406, Uttar Pradesh, India}

%%%%%%%%%%%%%%%%%%%%%%%%%%%%%%%%%%%%%  DATE  %%%%%%%%%%%%%%%%%%%%%%%%%%%%%%%%%%%%

\date{\today}

\begin{abstract}
Perturbative Quantum Chromodynamics (pQCD) corrections and color superconductivity predict that strongly interacting matter can reveal new physical phenomena under extreme conditions. Taking into account these interaction effects, we investigate the role of anisotropic pressure in quark stars composed of interacting quark matter. Adopting two physically well-motivated anisotropy profiles, we numerically solve the stellar structure equations in order to explore the consequences of anisotropic pressure on various macroscopic properties such as radius, gravitational mass, surface redshift, moment of inertia, tidal Love number and oscillation spectrum. Remarkably, for both anisotropy models, negative anisotropies increase the radial stability of interacting quark stars, while the opposite occurs for positive anisotropies. However, for the Bowers-Liang profile, the central density corresponding to the maximum-mass point does not coincide with the central density where the squared oscillation frequency vanishes, indicating that the existence of stable anisotropic interacting quark stars is possible beyond the maximum mass for negative anisotropies. Additionally, we compare our theoretical predictions with several observational mass-radius measurements and tidal deformability constraints, which suggest that both strong interaction effects and anisotropy effects play a crucial role in describing compact stars observed in the Universe.

\end{abstract}

\maketitle

\section{Introduction}

Einstein's theory of general relativity (GR) has stood like a pillar of modern theoretical physics regarding gravitational phenomena from the solar system to cosmological scales \cite{Will:2014kxa}. The success of this theory comes through the first experimental test by Arthur Eddington in 1919 during a total solar eclipse, i.e., the deflection of the starlight next to the Sun. Since then GR is the simplest metric theory and it remains to be the most successful gravity theory for understanding the Universe. Among many astonishing predictions of GR, compact astrophysical objects such as black holes, neutron stars (NSs), white dwarfs have turned from purely mathematical objects to potentially real physical entities as a result of different observational astronomical measurements.

In 1967, the discovery of pulsars had a great impact on astronomers in general. This discovery proves the existence of NSs in the Universe and is crucial to understand the nature of ultra-dense compact objects. NSs are the incredibly dense remnants of high mass stars when they run out of fuel. Once the core of the massive stars has completely burned to iron, the energy production stops and the core rapidly collapses, squeezing electrons and protons together to form neutrons and neutrinos. Such stars are supported by neutron-degeneracy pressure, and thus, are the most compact stars in the Universe. A NS that has a mass between $M \sim 1-3 M_{\odot}$ where $M_{\odot} =2\times 10^{33}\, \rm g$ with radius between $10-15\, \rm km$ \cite{Ozel:2016oaf,Steiner:2017vmg}. Consequently, their central densities are extremely high and easily exceed the nuclear saturation limit, i.e., $\rho \gtrsim \rho_{\text{nuc}} $ where $\rho_{\text{nuc}} = 2.8 \times 10^{14}$\, $\rm g/cm^3$. Indeed, it is very hard to deal with dense matter at such extreme situation in a laboratory conducted on Earth, and hence no comprehensive picture has been achieved till date.

Moreover, observed pulsars through electromagnetic (EM) signals have put a strong constraint on the equation of state (EoS) 
of dense matter in the interior of NSs. Meanwhile, the mass-radius measurements from spectroscopic observations of thermonuclear X-ray bursts, along with recent NICER (Neutron Star Interior Composition Explorer) data have significantly placed 
even tighter constraints on the EoS \cite{Miller:2019cac}. Consequently, physicists have conjectured the existence of more exotic states such as strange quark matter (SQM) in the core of compact objects. As a matter of fact, was first speculated that compact stars could be partially or totally made of SQM \cite{Itoh, Witten:1984rs, Bodmer:1971we}. It has been suggested that SQM consists of almost equal numbers of $u$, $d$ and $s$ quarks, and a small number of electrons to attain the charge neutrality. Quarks are strongly interacting particles and may exist from a few fermis up to a large (kilometer-sized) ranging in size with the possibility of of forming consistent self-bound quark stars (QSs). The simple model proposed for SQM is the MIT bag model \cite{Farhi:1984qu} in which the quarks are considered to be free inside a bag. Depending on this model, the internal structure of QSs has been explored by several authors in Einstein gravity \cite{Nicotra:2006eg, Arbanil:2016wud, Joshi:2020lwn, Lopes2021PS, Lopes:2020dvs, Arbanil2023} and modified gravity theories \cite{Astashenok2015, Salako:2021xkj, Pretel:2022plg, PretelTA2023}.

Most of the studies concerning the internal structure of NSs/QSs  is fully determined based on the assumption of isotropic matter,
where the stellar fluid is described by an isotropic perfect fluid. However, the discovery of more massive compact stars has turned out the situation more complex inside a compact stellar object, i.e., taking into consideration a transverse pressure in addition to a radial one \cite{Roupas2021}. Under this anisotropic picture, Bowers \& Liang \cite{Bowers:1974tgi} examined the possible effect of locally anisotropic EoS for relativistic spheres. They showed that anisotropy affects the maximum equilibrium mass and surface redshift of NSs. Indeed, it has been realized that local anisotropy in matter fields may arise for several reasons such as superfluid cores, strong magnetic fields, crystallization of the core or phase transitions, which play an important role in determining the microscopic composition of compact stars. But, the isotropy should be eventually restored if one considers an anisotropic matter source for compact stars.

With account of the anisotropy matter field, various solutions have been obtained in the literature \cite{Herrera:1994qua, Mak:2001eb, Lake:2009cd, Ivanov:2017kyr, Stelea:2018cgm, Biswas2019, DasCPC2023, Baskey2023, Parida2023}. It was argued that the quasi-local variables can construct an EoS to describe the anisotropic fluid in spherical symmetry \cite{Horvat:2010xf}. The effects of anisotropy on the global stellar properties of NSs such as their mass-radius relation, the moment of inertia and tidal deformability were studied in Refs.~\cite{Pretel:2020xuo, Curi:2022nnt, Das:2023pfq, Das:2023jtj}. Within an anisotropic context, the radial stability was rigorously addressed for quark stars \cite{Arbanil:2016wud}, neutron stars \cite{Pretel:2020xuo} and dark energy stars \cite{Pretel2023} through radial and adiabatic perturbations. Based on the Hartle-Thorne formalism, the mass correction and deformation of slowly rotating anisotropic NSs was discussed by Pattersons and Sulaksono \cite{Pattersons:2021lci}. See also Refs.~\cite{Silva:2014fca, Pretel:2022qng} for slowly rotating anisotropic NSs in modified gravity. In this work we will undertake how the anisotropic pressure affects the internal composition of compact stars with a more general and realistic equation of state for quark matter, namely, an interacting quark EoS.

The rest of the paper is organized as follows. In Sec.~\ref{sec2}, we preferred to describe the quark matter EoS which includes pQCD corrections and color superconductivity. In Sec.~\ref{sec3}, we present the gravitational field equations for an anisotropic fluid distribution assuming a  static, spherically symmetric geometry. The Tolman-Oppenheimer-Volkoff (TOV) equations are also presented with proper boundary conditions. In the same section, we study various macroscopic properties of anisotropic interacting quark stars (IQSs) such as moment of inertia, tidal Love number and radial vibration frequencies. In Sec.~\ref{sec4}, we employ two different anisotropy models for QSs in strong gravitational fields. In next, we continue our discussion for numerical findings specially focusing on the mass-radius relation and the radial stability with varying degrees of anisotropy in Section \ref{sec5}. We further compare our theoretical predictions with observational data in Sec.~\ref{sec6}. Finally, a summary is provided in Sec.~\ref{sec7}.

%%%%%%%%%%%%%%%%%%%%%%%%%%%%%%%%%%%%%%%%%%%%%%%%%%%%%%%%%%%%%%%%%%%
 \section{Interacting quark matter EoS } \label{sec2}
%%%%%%%%%%%%%%%%%%%%%%%%%%%%%%%%%%%%%%%%%%%%%%%%%%%%%%%%%%%%%%%%%%
Let us start by reviewing the main results contained in Ref. \cite{Zhang:2020jmb}, where authors have established a general parametrization of the quark matter EoS which is specified by the perturbative QCD (pQCD) correction and color superconductivity~\cite{Zhang:2020jmb}. One interesting feature is that depending only on a single parameter one can convert the EoS into a dimensionless form. Moreover, the rescaling process  reduces the number of degrees of freedom of the given EoS. In the current study, we are  searching for the presence of anisotropic matter, and how it affects the macroscopic properties of SQSs. 

To model the interacting quark matter, the relation between the radial pressure $(P_r)$ to the energy density $(\rho)$ is given by~\cite{Zhang:2020jmb, Zhang:2021fla}: 
\begin{align}\label{eos_tot}
  P_r=&\ \frac{1}{3}(\rho-4B_{\rm eff})  \nonumber  \\
  &+ \frac{4\lambda^2}{9\pi^2}\left[ -1+{\rm sgn}(\lambda)\sqrt{1+3\pi^2 \frac{(\rho-B_{\rm eff})}{\lambda^2}} \right], 
\end{align}
where the bag constant as an effective parameter is denoted by $B_{\rm eff}$ that accounts for the nonperturbative contribution from the QCD vacuum and
\begin{eqnarray}
\lambda=\frac{\xi_{2a} \Delta^2-\xi_{2b} m_s^2}{\sqrt{\xi_4 a_4}}.
\label{lam}
\end{eqnarray}

One can immediately check that when $\lambda \to 0$, the Eq. (\ref{eos_tot}) become the conventional noninteracting quark matter EoS $P_r= (\rho - 4B_{\rm eff})/3$. We can therefore say that the parameter $\lambda$ is a relative measure of the strong interaction effects. In the following, $\Delta$ denotes the gap parameter and $m_s$ represents the strange quark mass. The pQCD correction parameter $a_4$ can vary from small values to $a_4=1$, which gives a larger deviation from an ideal relativistic Fermi gas EoS. Here,  
${\rm sgn}(\lambda)$ represents the sign of $\lambda$, and leads to a positive value when 
$\Delta^2/m_s^2>\xi_{2b}/\xi_{2a}$. The constant coefficients in $\lambda$ are
\begin{align}
(\xi_4,\xi_{2a}, \xi_{2b}) = \left\{ \begin{array} {ll}
(( \left(\frac{1}{3}\right)^{\frac{4}{3}}+ \left(\frac{2}{3}\right)^{\frac{4}{3}})^{-3},1,0) & \textrm{2SC phase}\\
(3,1,3/4) & \textrm{2SC+s phase}\\
(3,3,3/4)&   \textrm{CFL phase}
\end{array}
\nonumber
\right.
\end{align}
characterizing the possible phases of color superconductivity. We can further introduce the dimensionless rescaling
\begin{align}
 \bar{\lambda}=\frac{\lambda^2}{4B_{\rm eff}}= \frac{(\xi_{2a} \Delta^2-\xi_{2b} m_s^2)^2}{4\,B_{\rm eff}\xi_4 a_4},
 \label{rescaling_lam}
\end{align}
and
\begin{align}
\bar{\rho} &=\frac{\rho}{4\,B_{\rm eff}},  & 
 \bar{P}_r &=\frac{P_r}{4\,B_{\rm eff}}.
\label{rescaling_prho}
\end{align}
Consequently, removing the $B_{\rm eff}$ parameter, Eq.~(\ref{eos_tot}) assumes the following dimensionless form 
\begin{align}\label{eos_p}
  \bar{P}_r =&\ \frac{1}{3}(\bar{\rho}-1) + \frac{4\bar{\lambda}}{9\pi^2} \left[-1+ \sqrt{1+ \frac{3\pi^2}{\bar{\lambda}} \left(\bar{\rho}-\frac{1}{4}\right)}\right] .
\end{align}

\begin{figure}
    \centering
    \includegraphics[width= 8.2cm]{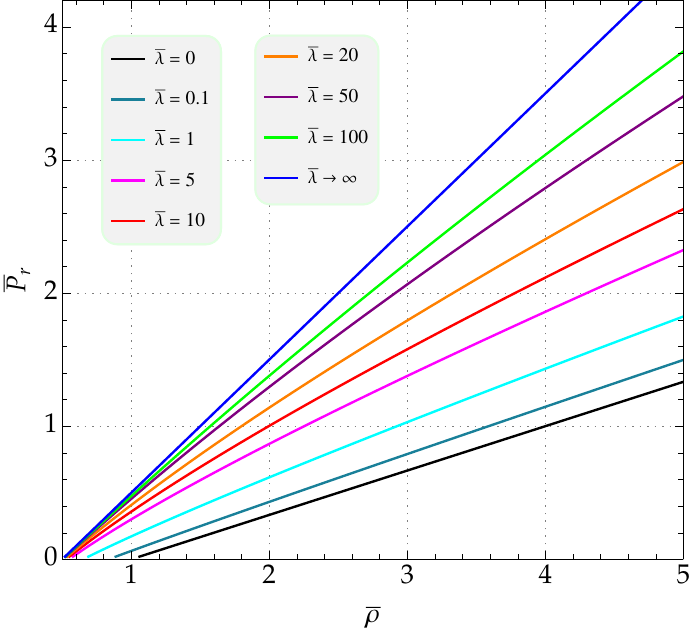}
    \caption{Equation of state involving interacting quark matter in its dimensionless form (\ref{eos_p}) for several values of $\bar{\lambda}$, where the linear behavior for $\bar{\lambda} \to \infty$ is described by the blue curve, see Eq.~(\ref{eos_infty1}). Moreover, we recover the standard noninteracting quark matter EoS when $\bar{\lambda} =0$, which is represented by the black curve.  }
    \label{FigIQMEoS}
\end{figure}

\begin{figure}
    \centering
    \includegraphics[width= 8.5cm]{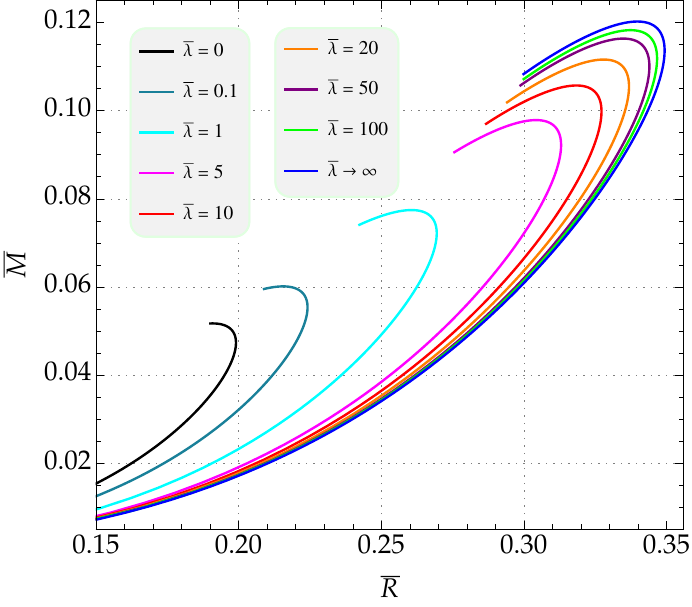}
    \caption{Mass-radius diagram of interacting QSs with EoS (\ref{eos_p}) for various values of $\bar{\lambda}$ as shown in Fig.~\ref{FigIQMEoS}. Here, the radius and mass have been rescaled into a dimensionless form as $\bar{R}= R\sqrt{4B_{\rm eff}}$ and $\bar{M}= M\sqrt{4B_{\rm eff}}$, respectively, see Ref.~\cite{Zhang:2021fla} for further details. A stiffer EoS, which is achieved for large values of $\bar{\lambda}$, is capable of generating more massive QSs. }
    \label{FigMRDiagram}
\end{figure}

It can be seen that for extremely large positive values of  $\bar{\lambda}$, Eq.~(\ref{eos_p}) approaches the special form 
\begin{align}
\bar{P}_r\vert_{\bar{\lambda}\to \infty}=\bar{\rho}-\frac{1}{2}, 
\label{eos_infty1}
\end{align}
which is equivalent to $P_r = \rho- 2B_{\rm eff}$. In Fig.~\ref{FigIQMEoS} we show this linear behavior for a very large $\bar{\lambda}$ by a blue curve. We further note that the parameter $\bar{\lambda}$ gives a stiffer EoS for positive increasing values of $\bar{\lambda}$, where the well known case $\bar{\lambda} =0$ describes ordinary noninteracting quark matter. With these increasing values one can reach sufficiently high masses \cite{Zhang:2020jmb, Zhang:2021fla} for IQSs. This can be clearly observed in the mass-radius diagram displayed in Fig.~\ref{FigMRDiagram} into a dimensionless form. The advantage of using a dimensionless description for the different variables is that it becomes unnecessary to specify the value of the effective bag constant, unless we want to obtain the global characteristics of quark stars in physical units for comparison reasons. Furthermore, it was shown by Zhang \cite{Zhang:2021fla} that it is possible to produce gravitational wave echoes for IQSs when $\bar{\lambda} \gtrsim 10$ at large central pressure. In the present work, we will consider a value of $\bar{\lambda}$ for which the strong interaction effects are relatively appreciable, e.g.~see the red curve in Fig.~\ref{FigIQMEoS}. However, in section \ref{sec6} we will adopt smaller values in order to compare our results with observational data. Of course, the qualitative behavior of the results must hold as $\bar{\lambda}$ varies.

%%%%%%%%%%%%%%%%%%%%%%%%%%%%%%%%%%%%%%%%%%%%%%%%%%%%%%
\section{Equilibrium anisotropic configurations} \label{sec3}
%%%%%%%%%%%%%%%%%%%%%%%%%%%%%%%%%%%%%%%%%%%%%%%%%%%%%

\subsection{TOV equations}

In the present work we assume an anisotropic matter distribution where the radial pressure $P_r$ differs from the transverse pressure $P_{\perp}$. The covariant form of the energy-momentum tensor of such source may be written as 
\begin{eqnarray}\label{EMT}
T_{\mu \nu} = (\rho + P_{\perp})u_\mu u_\nu + P_{\perp}g_{\mu \nu} - \sigma \chi_{\mu} \chi_{\nu}, 
\end{eqnarray}
where $u^{\mu}$ is the 4-velocity for a comoving fluid element, $\chi^{\mu}$ is the unit spacelike vector in the radial direction and $g_{\mu\nu}$ is the metric tensor. Moreover, $\rho$ is the energy density while $\sigma\equiv P_{\perp}- P_{r}$ is the anisotropic factor in the source, with $\sigma= 0$ corresponding to isotropic matter. The four-vectors $u^{\mu}$ and $\chi^{\mu}$ must satisfy the following relations 
\begin{align}
u_{\mu}u^{\mu} &=-1,  &   \chi_{\mu}\chi^{\mu} &=1,  &   u_{\mu}\chi^{\mu} &=0.
\end{align}

Considering static and spherically symmetric compact stars, we use the usual line element given by 
\begin{eqnarray}\label{metric}
  ds^2 = - e^{2\Phi(r)}dt^2 + e^{2\Psi(r)}dr^2 + r^2(d\theta^2 + \sin^2\theta d\phi^2 ),
\end{eqnarray}
where $\Phi(r)$ and $\Psi(r)$ are unknown metric potentials to be determined. The gravitational mass enclosed within a sphere of radius $r$, denoted by $m(r)$, is related to the metric function $\Psi(r)$ through $e^{-2\Psi} = 1- 2m/r$. Consequently, the  Einstein field equation, $R_{\mu\nu} - g_{\mu\nu}R/2 = 8\pi T_{\mu\nu}$ (we choose units with $G = c = 1$), provides the following stellar structure equations:
\begin{eqnarray}
  && m^{\prime}= 4\pi r^2 \rho , \label{tov1}\\
  && P^{\,\prime}_r= - (\rho+ P_r)\, \frac{m+ 4\pi r^3 P_r }{r\left(r - 2m\right)} + \frac{2\sigma}{r}, \label{tov2} \\
  && \Phi'= -\frac{P_r'}{\rho+ P_r} + \frac{2\sigma}{r(\rho+ P_r)} \label{tov3} ,
\end{eqnarray}
where the prime denotes a derivative with respect to $r$.

This set of equations (\ref{tov1})-(\ref{tov3}) with the EoS (\ref{eos_tot}) represents the internal structure of QSs. To solve them numerically we have to choose a specific value for $\rho_c$ (central energy density) with the following natural boundary conditions:
\begin{align}\label{eq:r0}
  \rho(0) &= \rho_c, &   m(0) &= 0,
\end{align}
and then integrate towards the surface of the star where the pressure drops to zero. This is identified as the stellar radius at $P_r(R)=0$. Additionally, boundary conditions are required to match this interior solution to an external metric, which is the Schwarzschild exterior solution. This is defined by
\begin{equation}
e^{2\Phi(R)} = 1-\frac{2M}{R}  
\end{equation}
with $M = m(r=R)$ being the total mass of the star. Consequently, the surface gravitational redshift can be written in terms of radius and mass as follows
\begin{equation}\label{RedshiftEq}
    z_{\rm sur} = \left[ 1- \frac{2M}{R} \right]^{-1/2} - 1 .
\end{equation}

\subsection{Moment of inertia}

Under the slowly rotating approximation (i.e., we consider only first-order terms in the angular velocity of the star $\Omega$) \cite{Hartle1967}, the moment of inertia of an anisotropic compact star is given by \cite{Pretel2023}
\begin{equation}\label{MomInerEq}
    I = \frac{8\pi}{3}\int_0^R (\rho+ P_r+ \sigma)e^{\Psi-\Phi}r^4 \left( \frac{\varpi}{\Omega} \right)dr ,
\end{equation}
where $\varpi = \Omega- \omega$, with $\omega(r,\theta)$ being the angular velocity acquired by an observer falling freely from infinity. In fact, it is possible to show that $\varpi$ is a function only of the radial coordinate and is determined after solving the following differential equation (see Ref.~\cite{Pretel2023} for further details)
\begin{equation}\label{OmegaEq}
    \frac{e^{\Phi-\Psi}}{r^4}\frac{d}{dr}\left[ e^{-(\Phi+\Psi)}r^4\frac{d\varpi}{dr} \right] = 16\pi(\rho+ P_r+ \sigma)\varpi .
\end{equation}

Suitable boundary conditions for the above differential equation come from the requirements of regularity at the origin and asymptotic flatness at very far distances from the star. Note that, at infinity (where spacetime is flat) the dragging angular velocity $\omega(r)$ goes to zero. Therefore, for an arbitrary choice of $\varpi(0)$, Eq.~(\ref{OmegaEq}) has to satisfy
\begin{align}\label{BCMomIner}
    \left. \frac{d\varpi}{dr}\right\vert_{r=0} &= 0,  &  \lim_{r\rightarrow\infty}\varpi &= \Omega .
\end{align}

\subsection{Tidal Love number}

Neutron stars are tidally deformed under the influence of an external tidal field (created by a companion star) \cite{Chatziioannou2020}. This deformation, quantified through a parameter termed the tidal deformability $\Lambda$, can be measured in gravitational waves emitted from the inspiral of a binary neutron-star coalescence. In that regard, here we calculate the tidal deformability of individual anisotropic QSs. The dimensionless tidal deformability is determined from the compactness $C= M/R$ and the tidal Love number $k_2$ via \cite{Kanakis2020, Yang2023, Arbanil2023}
\begin{equation}\label{TidalDeforEq}
    \Lambda = \frac{2}{3}\frac{k_2}{C^5} ,
\end{equation}
with $k_2$ being given by the following expression
\begin{align}\label{LoveNumEq}
    k_2 &= \frac{8}{5}(1- 2C)^2C^5 \left[ 2C(\alpha -1) - \alpha+ 2 \right]  \nonumber  \\
    &\times \left\lbrace 2C[ 4(\alpha+ 1)C^4 + (6\alpha- 4)C^3 \right.  \nonumber  \\
    &\left.+\ (26- 22\alpha)C^2 + 3(5\alpha -8)C - 3\alpha+ 6 \right]   \nonumber  \\
    &\left.+\ 3(1-2C)^2\left[ 2C(\alpha- 1)- \alpha +2 \right]\ln(1-2C) \right\rbrace^{-1} ,
\end{align}
where $\alpha= y(R)- 4\pi R^3 \rho_s/M$. Here, $\rho_s$ stands for the energy density at the surface of the interacting QS. The correction term ``$-4\pi R^3 \rho_s/M$'' is included because the energy density at the surface is finite and non-null \cite{Yang2023, Arbanil2023}. Actually, this becomes evident from the EoS (\ref{eos_tot}).

The function $y(r)$ is obtained by solving the differential equation:
\begin{equation}\label{yEq}
    ry' = -y^2 + (1 - r\mathcal{P})y - r^2\mathcal{Q} ,
\end{equation}
where we have defined \cite{Arbanil2023}
\begin{align}
    \mathcal{P} =&\ \frac{2}{r} + e^{2\Psi}\left[ \frac{2m}{r^2} + 4\pi r(P_r - \rho) \right] ,  \\
    \mathcal{Q} =&\ 4\pi e^{2\Psi}\left[ 4\rho + 4P_r+ 4P_{\perp} + \frac{\rho+ P_r}{\mathcal{A}v_{sr}^2}(1+ v_{sr}^2) \right]  \nonumber  \\
    &- \frac{6e^{2\Psi}}{r^2} - 4\Phi'^2 ,
\end{align}
with $\mathcal{A} = dP_{\perp}/dP_r$ and $v_{sr}= \sqrt{dP_r/d\rho}$ being the radial speed of sound. Besides, the boundary condition for Eq.~(\ref{yEq}) at the stellar origin is $y(0)=2$ \cite{Postnikov2021}. Therefore, $y(R)$ is calculated after integrating equation (\ref{yEq}) from the origin up to the surface of the anisotropic compact star. With that we determine $\alpha$, and consequently the tidal Love number (\ref{LoveNumEq}).

\subsection{Radial stability}

The TOV equations provide equilibrium solutions, however, such an equilibrium state can be stable or unstable \cite{Chandrasekhar1964}. To examine the stellar stability of the equilibrium anisotropic configurations with interacting quark matter, it becomes necessary to determine the oscillation frequencies when a star is radially and adiabatically perturbed. The radial vibrations in the interior of an anisotropic compact star are governed by \cite{Pretel:2020xuo}
\begin{align}
  \frac{d\zeta}{dr} =& -\frac{1}{r}\left( 3\zeta + \frac{\Delta P_r}{\gamma P_r} + \frac{2\sigma\zeta}{\rho + P_r} \right) + \frac{d\Phi}{dr}\zeta ,  \label{ROEq1}   \\
  \frac{d(\Delta P_r)}{dr} =&\ \zeta\left\lbrace \nu^2e^{2(\Psi - \Phi)}(\rho + P_r)r - 4\frac{dP_r}{dr} \right.  \nonumber  \\
  &\left. - 8\pi (\rho + P_r)e^{2\Psi}rP_r + r(\rho + P_r)\left(\frac{d\Phi}{dr}\right)^2 \right.  \nonumber \\
  &\left. + 2\sigma\left(\frac{4}{r} + \frac{d\Phi}{dr} \right) + 2\frac{d\sigma}{dr} \right\rbrace + 2\sigma\frac{d\zeta}{dr}   \nonumber \\
  & - \Delta P_r \left[ \frac{d\Phi}{dr} + 4\pi(\rho + P_r)re^{2\Psi} \right] + \frac{2}{r}\delta\sigma ,  \label{ROEq2} 
\end{align}
where the perturbation $\zeta$ is related to the Lagrangian displacement $\xi$ via $\zeta= \xi/r$. This system of first-order differential equations has been obtained assuming that the perturbations have a harmonic time dependence of the form $\sim e^{i\nu t}$, where $\nu$ is the oscillation frequency to be determined. Moreover, $\Delta P_r$ is the Lagrangian perturbation of the radial pressure, $\delta\sigma$ is the Eulerian perturbation for the anisotropy factor, and $\gamma= (1+\rho/P_r)dP_r/d\rho$ is the adiabatic index. Notice that, for any metric or fluid variable $f$, the relation between the Eulerian and Lagrangian perturbations is given by $\Delta f= \delta f + f'\xi$. The expression for $\delta\sigma$ will depend on the specific form of the anisotropic profile.

At the surface of the star, where the pressure is zero, the Lagrangian perturbation of the radial pressure must satisfy $\Delta P_r = 0$. Meanwhile, the following boundary condition must be imposed at the stellar center
\begin{equation}\label{BCRO1}
    \Delta P_r = -\frac{2\sigma\zeta}{\rho + P_r}\gamma P_r -3\gamma\zeta P_r  \qquad \  {\rm as}  \qquad \  r\rightarrow 0 .
\end{equation}

\begin{figure*}
    \centering
    \includegraphics[width= 5.49cm]{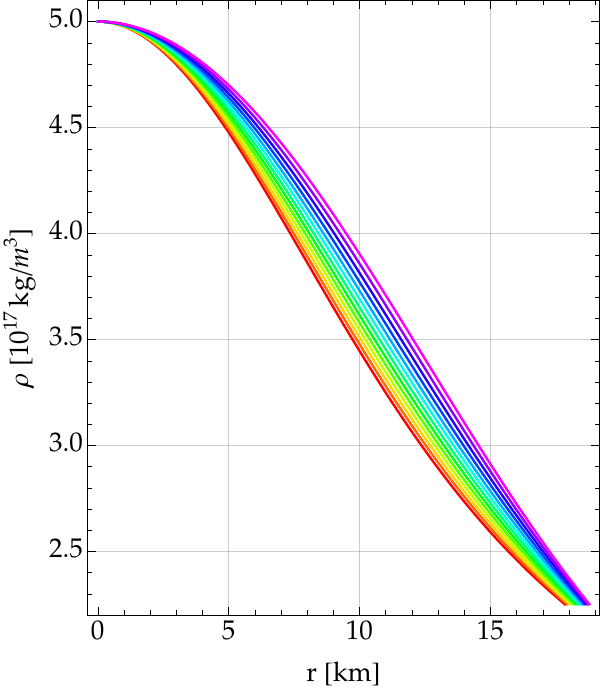}
    \includegraphics[width= 5.25cm]{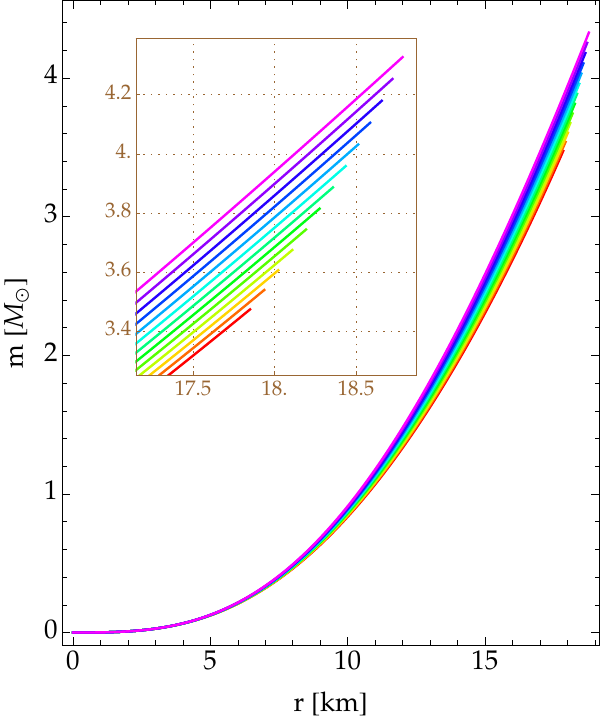}
    \includegraphics[width= 6.69cm]{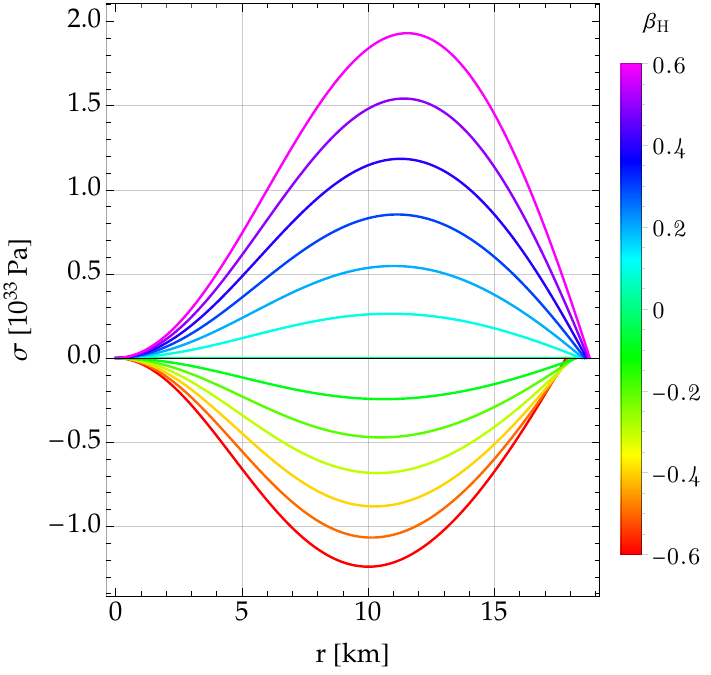}
    \caption{Radial profiles of the energy density (left panel), mass function (middle panel) and anisotropy factor (right panel) inside an anisotropic interacting quark star with central density $\rho_c= 0.5 \times 10^{18}\, \rm kg/m^3$ and EoS (\ref{eos_tot}) for $B_{\rm eff}= 60\, \rm MeV/fm^3$ and $\Bar{\lambda}= 10$. Here we have used the quasi-local model (\ref{aniso_mod1}) for values of $\beta_{\rm H}$ in the range $\beta_{\rm H} \in [-0.6, 0.6]$, see the color bar on the right side of the third plot. The main consequence of positive (negative) anisotropy is a relevant increase (decrease) in the radius and mass of the star. Remark that the anisotropy is more pronounced in the intermediate regions, and the particular case $\beta_{\rm H}= 0$ corresponds to zero anisotropic pressure. }
    \label{FigRadialProfilesH}
\end{figure*}

\begin{figure*}
    \centering
    \includegraphics[width= 5.49cm]{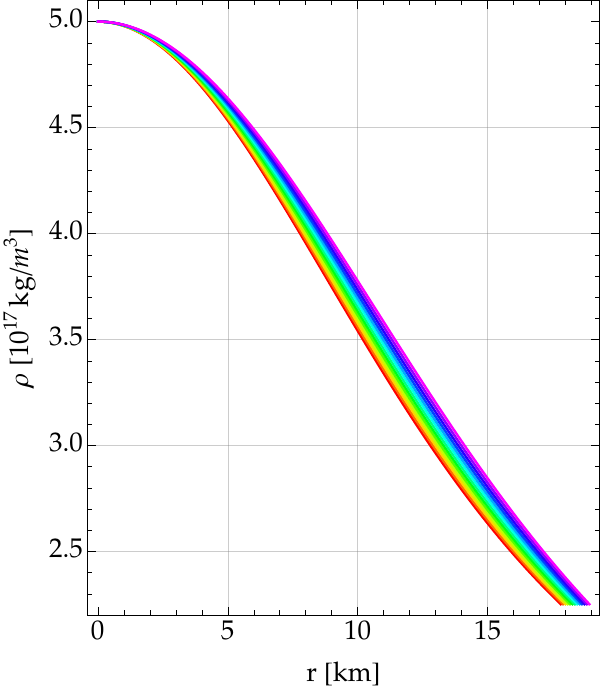}
    \includegraphics[width= 5.25cm]{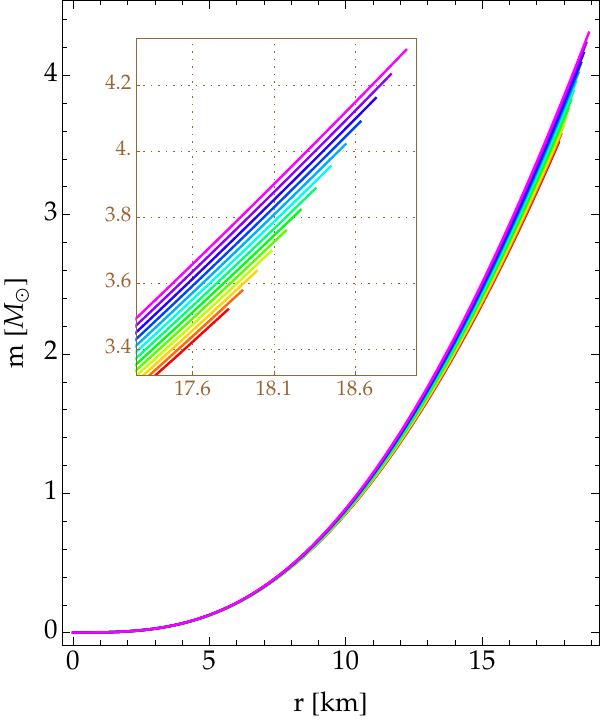}
    \includegraphics[width= 6.69cm]{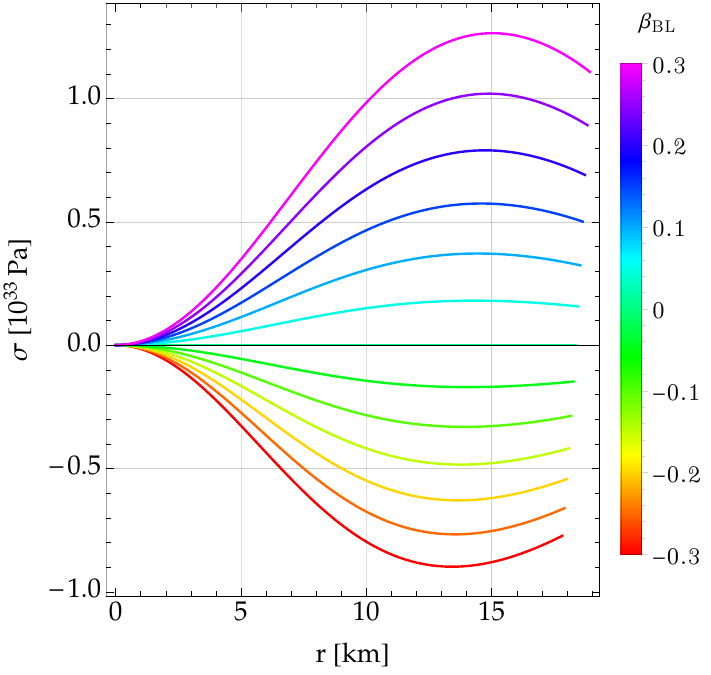}
    \caption{Energy density, mass and anisotropy as functions of the radial coordinate $r$, as in Fig.~\ref{FigRadialProfilesH}, but for the Bowers-Liang model (\ref{aniso_mod2}) with anisotropy parameter $\beta_{\rm BL} \in [-0.3, 0.3]$. The radial behavior for $\rho$ and $m$ is similar to those generated by the quasi-local model, however, the anisotropy is not zero at the surface of the star. }
    \label{FigRadialProfilesBL}
\end{figure*}

\begin{table*}
\caption{\label{table1} 
Global properties of interacting QSs with central density $\rho_c= 0.5 \times 10^{18}\, \rm kg/m^3$ and EoS (\ref{eos_tot}) for $B_{\rm eff}= 60\, \rm MeV/fm^3$ and $\Bar{\lambda}= 10$. The radial behavior of the energy density, mass and anisotropy factor of these stars is shown in Figs.~\ref{FigRadialProfilesH} and \ref{FigRadialProfilesBL} for the quasi-local ansatz and Bowers-Liang profile, respectively. Note that the frequency of the fundamental mode and first overtone are given by $f_0= \nu_0/2\pi$ and $f_1= \nu_1/2\pi$, respectively. }
\begin{ruledtabular}
\begin{tabular}{cccccccc}
Free parameter  &  $R [\rm{km}]$  &  $M [M_\odot]$  &  $z_{\rm sur}$   &   $I [10^{39}\, \rm kg \cdot m^2]$  &   $k_2$   &   $f_0 [\rm kHz]$   &   $f_1 [\rm kHz]$  \\
\colrule
Isotropic case  &  18.358  &  3.886  &  0.633  &  1.441  &  0.040  &  1.323  &  4.787  \\
$\beta_{\rm H}= -0.6$  &  17.850  &  3.473  &  0.533  &  1.157  &  0.061  &  1.586  &  5.576  \\
$\beta_{\rm H}= -0.3$  &  18.109  &  3.674  &  0.579  &  1.291  &  0.050  &  1.462  &  5.186  \\

$\beta_{\rm H}= 0.3$  &  18.587  &  4.105  &  0.696  &  1.608  &  0.032  &  1.168  &  4.381  \\
$\beta_{\rm H}= 0.6$  &  18.785  &  4.327  &  0.768  &  1.788  &  0.025  &  0.993  &  3.971  \\
$\beta_{\rm BL}= -0.30$  &  17.818  &  3.519  &  0.549  &  1.170  &  0.052  &  1.722  &  5.543  \\
$\beta_{\rm BL}= -0.15$  &  18.085  &  3.696  &  0.588  &  1.297  &  0.046  &  1.527  &  5.169  \\
$\beta_{\rm BL}= 0.15$  &  18.636  &  4.089  &  0.685  &  1.607  &  0.034  &  1.108  &  4.398  \\
$\beta_{\rm BL}= 0.30$  &  18.916  &  4.305  &  0.747  &  1.798  &  0.029  &  0.875  &  4.002  \\
\end{tabular}
\end{ruledtabular}
\end{table*}

%%%%%%%%%%%%%%%%%%%%%%%%%%%%%%%%%%%%%%%%%%%%%%%%%%%%%%
\section{Overview of different anisotropy models}   \label{sec4}
%%%%%%%%%%%%%%%%%%%%%%%%%%%%%%%%%%%%%%%%%%%%%%%%%%%%%

To generate numerical solutions in the present work, we will adopt two different anisotropy models, which have been widely used in the literature to model anisotropic matter at high densities and pressures:

\begin{enumerate}
    \item The well-known Quasi-Local (QL) model as proposed by Horvat \textit{et al.}~\cite{Horvat:2010xf}, and define by 
    \begin{eqnarray}\label{aniso_mod1}
\sigma_1 &\equiv& P_{\perp} - P_r =  \beta_{\rm{H}} P_r \mu,
\end{eqnarray}
where the free parameter $\beta_{\rm{H}}$ measures the degree of anisotropy and the function $\mu= 2m(r)/r$ define the compactness. We focus in the range of $-2\leq \beta_{\rm{H}} \leq 2$, and has been deeply discussed in \cite{Silva:2014fca, Folomeev:2018ioy, Doneva:2012rd, Yagi:2015hda, Pretel:2020xuo, Pretel:2022qng, Rahmansyah:2020gar, Rahmansyah:2021gzt}.

The Eulerian perturbation of the anisotropy factor takes the form \cite{Pretel:2020xuo}
\begin{equation}
    \delta\sigma_1 = \beta_{\rm H}\left[ (1- e^{-2\Psi})\delta P_r + 2P_r e^{-2\Psi}\delta\Psi \right] ,
\end{equation}
where
\begin{equation}
    \delta\Psi= -4\pi r^2\zeta(\rho+ P_r)e^{2\Psi} .
\end{equation}

    \item We are also interested in the anisotropy factor suggested by Bowers and Liang \cite{Bowers:1974tgi}
\begin{eqnarray}
    \sigma_2 &\equiv& P_{\perp} - P_r = \beta_{\rm{BL}} (\rho + P_r) (\rho + 3 P_r) r^2 e^{2\Psi}, \label{aniso_mod2}
\end{eqnarray}
where the free parameter $\beta_{\rm{BL}}$ is an important quantity in describing the measure of anisotropy. Note that when $\beta_{\rm{BL}} \to 0$, the anisotropy pressure vanishes at the center and this guarantees the regularity of Eq.~(\ref{tov2}). The  meaningful value of $\beta_{\rm{BL}}$ lies in the range of $-2 \leq \beta_{\rm{BL}} \leq 2$, see Refs.~\cite{Biswas2019, Silva:2014fca}. 

For this model, the Eulerian perturbation of the anisotropy factor is given by \cite{Pretel:2020xuo}
\begin{align}
    \delta\sigma_2 =&\ 2\beta_{\rm BL}r^2e^{2\Psi}\left[ (\rho+P_r)(\rho+ 3P_r)(\zeta+ \delta\Psi)  \right.  \nonumber  \\
    &\left. +(2\rho+ 3P_r)\delta P_r + (\rho+ 2P_r)\delta\rho \right] ,
\end{align}
where
\begin{align}
    \delta\rho =& -2\zeta(\rho+P_r) - r\zeta(\rho'+ P_r') 
 \nonumber   \\
 &- (\rho+ P_r)(\zeta+ r\zeta') .
\end{align}

\end{enumerate}

To describe realistic astrophysical objects, certain acceptability conditions have to be established or obeyed at any spacetime point of the star \cite{Mak2003}: $(a)$ Energy density, radial pressure and tangential pressure should be positive, $(b)$ gradients for energy density and radial pressure must be negative, $(c)$ the radial and tangential speed of sound should be less than the speed of light, $(d)$ the interior solution must satisfy the energy conditions, $(e)$ isotropy at the stellar center, among others. As we will see in our results, the models adopted in this work satisfy such physical acceptability conditions.

Using the Bowers-Liang model (\ref{aniso_mod2}), Biswas and Bose showed that the anisotropic pressure in NSs can reduce their tidal deformability \cite{Biswas2019}. Additionally, they claimed that certain equations of state that are ruled out by GW170817 observations (using isotropic matter) can become viable if the stars had a significant enough anisotropic pressure component. Rahmansyah and Sulaksono \cite{Rahmansyah:2021gzt} have argued that the above two models are consistent with the recent NS multimessenger observations. Therefore, based on these interesting findings, there is a need to explore the astrophysical effect of each model on the internal structure of IQSs.

%%%%%%%%%%%%%%%%%%%%%%%%%%%%%%FIGURES%%%%%%%%%%%%%%%%%%%%%%%%%%%%

\section{Numerical results and discussion}\label{sec5}

Given an anisotropy profile for $\sigma$ and an EoS, we begin our analysis by solving the set of stellar structure equations (\ref{tov1})-(\ref{tov3}). For our numerical calculations we will use $B_{\rm eff}= 60\, \rm MeV/fm^3$ and $\Bar{\lambda}= 10$ in the equation of state (\ref{eos_tot}). Figure \ref{FigRadialProfilesH} illustrates the numerical solutions of the TOV equations for the anisotropy profile proposed by Horvat \textit{et al.}~\cite{Horvat:2010xf} by considering a central density $\rho_c= 0.5 \times 10^{18}\, \rm kg/m^3$ and anisotropy parameter varying in the range $\beta_{\rm H} \in [-0.6, 0.6]$. One observes that positive (negative) values of $\beta_{\rm H}$ increase (decrease) the mass and radius of the interacting QS. Furthermore, the anisotropy factor $\sigma$ vanishes at the origin (because of $\mu =0$ at $r=0$) and at the surface (since $P_r= 0$ at $r=R$), while the largest effect takes place in the intermediate zones. As is evident, the isotropic solutions are recovered when $\beta_{\rm H}= 0$. In Fig.~\ref{FigRadialProfilesBL} we present our numerical solutions for the anisotropy profile (\ref{aniso_mod2}) with $\beta_{\rm BL} \in [-0.3, 0.3]$. The radial behavior for the energy density and mass function is similar to the first model, however, the anisotropy factor $\sigma$ is no longer zero at the surface. This last characteristic is due to the fact that the energy density does not disappear at the surface of a QS even if the radial pressure is zero, see EoS (\ref{eos_tot}). Explicit values for the mass and radius of some stellar configurations represented in Figs.~\ref{FigRadialProfilesH} and \ref{FigRadialProfilesBL} are shown in Table \ref{table1}. It is worth mentioning that positive (negative) values of $\beta_{\rm H}$ and $\beta_{\rm BL}$ increase (decrease) the energy density and, according to Eq.~(\ref{tov1}), this should lead to increasing (decreasing) mass at any radial coordinate $r$ within the stellar fluid. Consequently, all global properties of an interacting quark star will be altered due to the change in energy density and mass function.

\begin{figure*}
    \centering
    \includegraphics[width= 8.8cm]{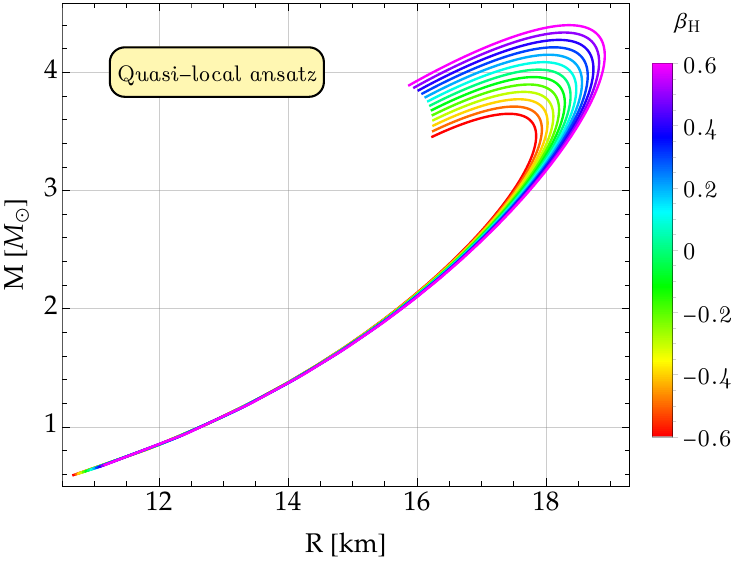}
    \includegraphics[width= 8.8cm]{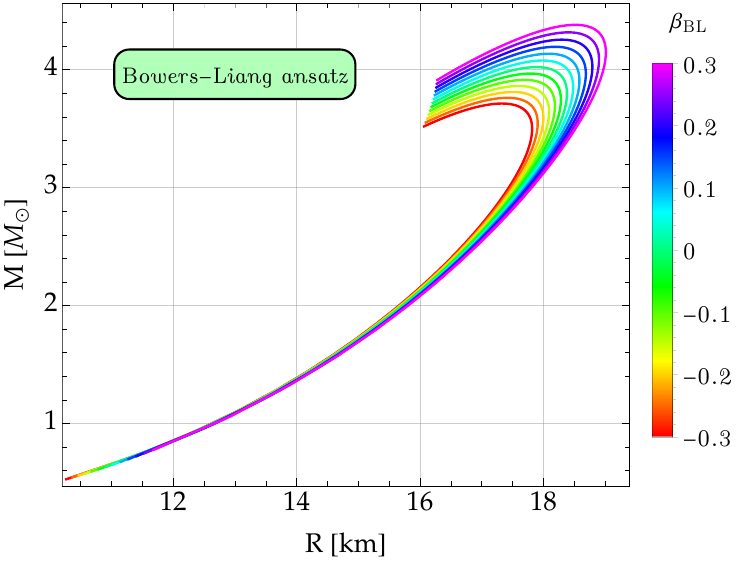}
    \caption{Mass-radius relations of the QSs with interacting quark matter EoS from Eq.~(\ref{eos_tot}) for $B_{\rm eff}= 60\, \rm MeV/fm^3$ and $\Bar{\lambda}= 10$. The left and right panel correspond to the anisotropic model I (\ref{aniso_mod1}) and II (\ref{aniso_mod2}), respectively. For the quasi-local ansatz, the anisotropy parameter $\beta_{\rm H}$ has been varied in the range $\beta_{\rm H} \in [-0.6,0.6]$. Meanwhile, for the model proposed by Bowers and Liang, we have assumed $\beta_{\rm BL} \in [-0.3, 0.3]$. }
    \label{FigMassRadius}
\end{figure*}

\begin{figure*}
    \centering
    \includegraphics[width= 8.8cm]{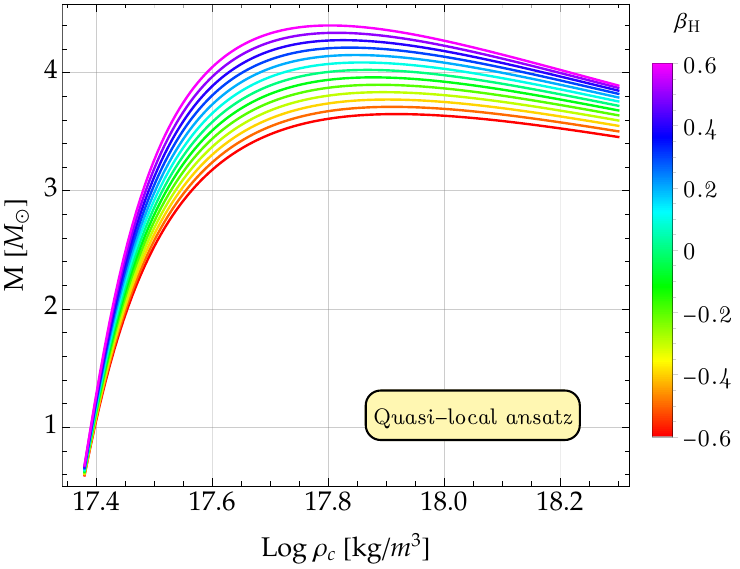}
    \includegraphics[width= 8.8cm]{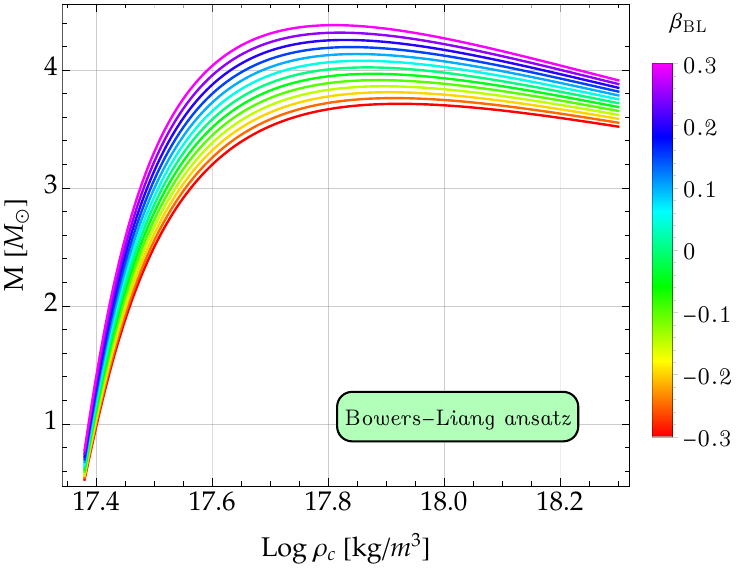}
    \caption{Gravitational mass as a function of the energy density corresponding to the equilibrium configurations shown in Fig.~\ref{FigMassRadius}. The increase in either of the parameters $\beta_{\rm H}$ or $\beta_{\rm BL}$ implies that the critical configuration (corresponding to the maximum-mass point) is found at a smaller central-density value. For the horizontal axis in both plots, we have used the common logarithm (i.e., the logarithm with base $10$). The same applies to Figs.~\ref{FigOmega2dHorvat} and \ref{FigOmega2dBL}.} 
    \label{FigMassCentDen}
\end{figure*}

\begin{figure*}
    \centering
    \includegraphics[width= 8.8cm]{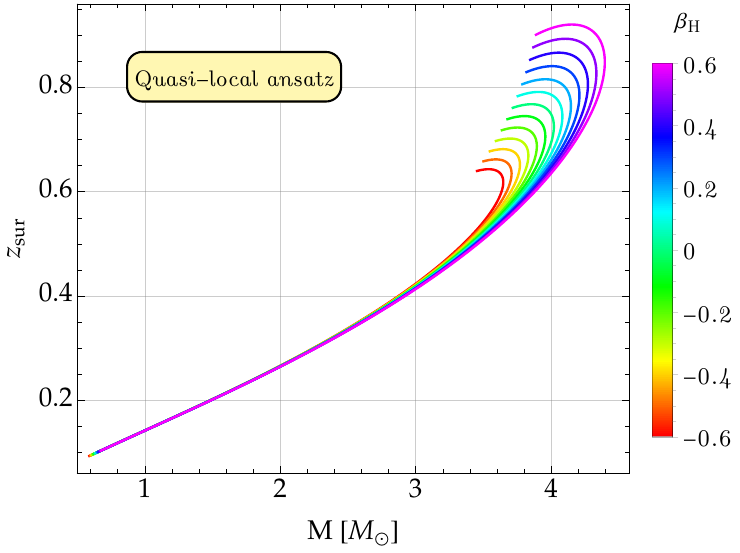}
    \includegraphics[width= 8.8cm]{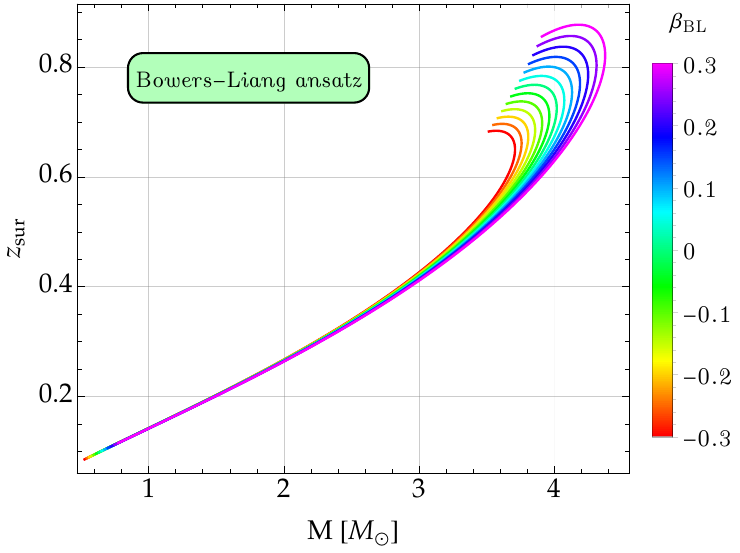}
    \caption{Surface gravitational redshift versus total mass for both anisotropy profiles. In both plots, the impact of anisotropy is substantial only in the high-mass region. Larger redshifts are achieved for positive anisotropies. }
    \label{FigRedsMass}
\end{figure*}

The variation of the central density $\rho_c$ leads us to obtain a family of QSs in hydrostatic equilibrium, which can be presented in the so-called mass-radius diagram. In Fig.~\ref{FigMassRadius}, we show such a diagram for a set of values of $\beta _{\rm H}$ (see left plot) and $\beta _{\rm BL}$ (see right plot). For both anisotropy models, we see that the main consequence of a positive anisotropy is a relevant increase in the maximum-mass values, while the opposite occurs for a negative anisotropy. The mass versus central density relations are displayed in Fig.~\ref{FigMassCentDen}, where it can be observed that the critical central density (corresponding to the maximum mass) is found to be at a lower value as the anisotropy increases. This means that, according to the standard criterion for stability $dM/d\rho_c >0$, a positive (negative) anisotropy decreases (increases) the stability of an interacting QS. Notwithstanding, this classical criterion is a necessary but not sufficient condition for stellar stability. A sufficient condition involves determining the oscillation frequencies when a compact star is perturbed. We will return to this later. Once the mass and radius are known, we can then determine the gravitational redshift via expression (\ref{RedshiftEq}). According to Fig.~\ref{FigRedsMass}, $z_{\rm sur}$ is affected by anisotropy mainly in the high-mass branch, while the changes are irrelevant at small masses. This behavior was already expected due to the mass-radius relations shown in Fig.~\ref{FigMassRadius}.

\begin{figure*}
    \centering
    \includegraphics[width= 8.5cm]{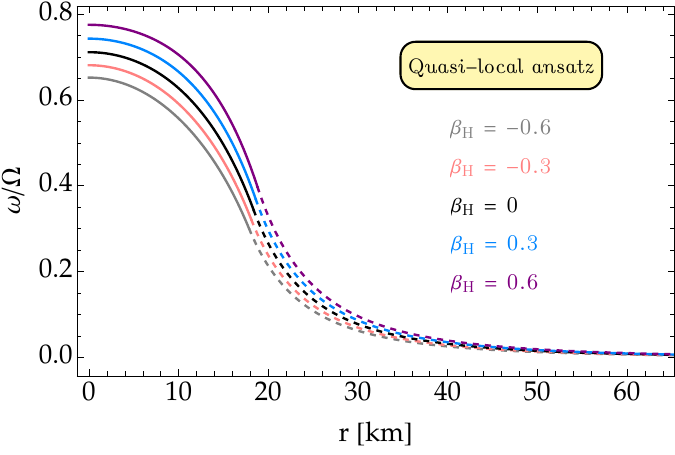}
    \includegraphics[width= 8.5cm]{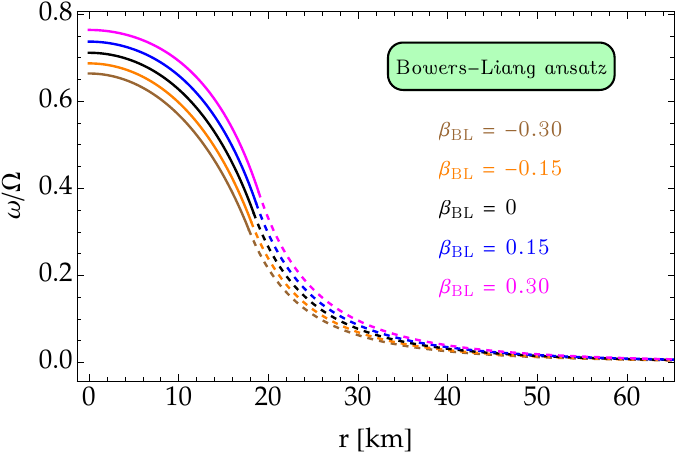}
    \caption{Frame-dragging angular velocity as a result of integrating the differential equation (\ref{OmegaEq}) for a central density $\rho_c= 0.5 \times 10^{18}\, \rm kg/m^3$ and EoS (\ref{eos_tot}) for $B_{\rm eff}= 60\, \rm MeV/fm^3$ and $\Bar{\lambda}= 10$. We have also considered five values for $\beta_{\rm H}$ and $\beta_{\rm BL}$. The solid lines represent the interior solutions, while dashed lines indicate the exterior region. A positive anisotropy leads to significantly increasing the angular velocity mainly in the innermost regions of the interacting quark star. As expected, the drag experienced (in the direction of rotation) by a particle falling toward the star is increasingly greater as it approaches the surface, so that $\omega\rightarrow 0$ for $r\rightarrow\infty$.  }
    \label{FigAngVelo}
\end{figure*}

\begin{figure*}
    \centering
    \includegraphics[width= 8.8cm]{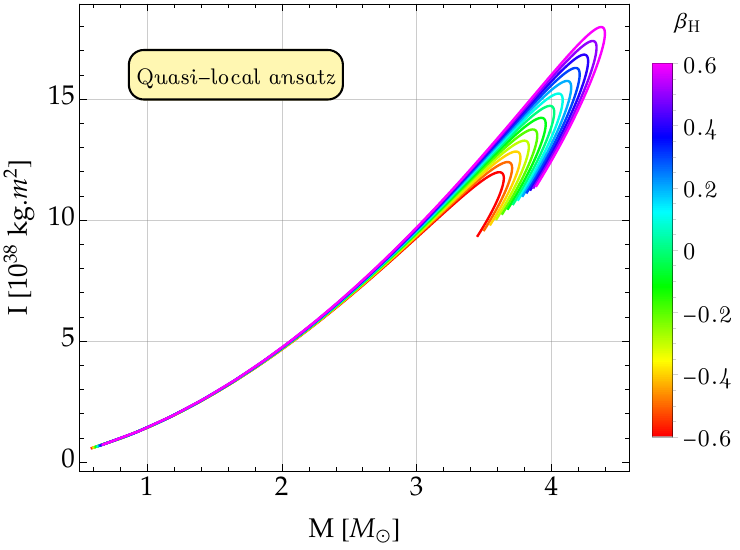}
    \includegraphics[width= 8.8cm]{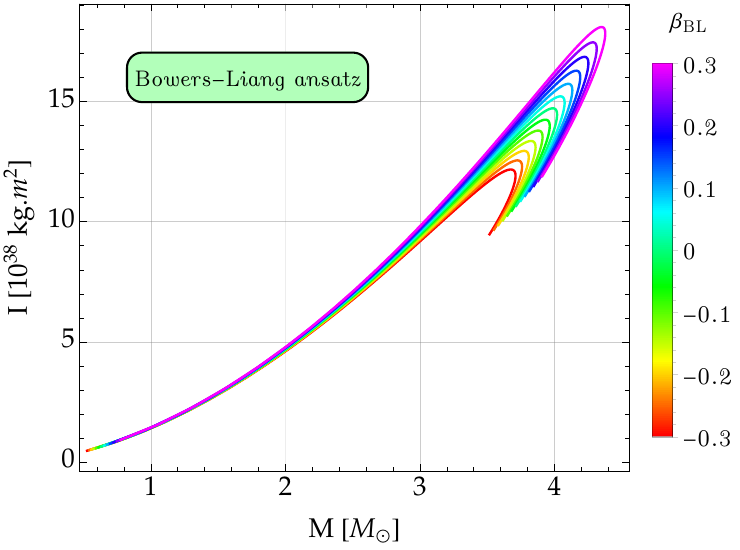}
    \caption{Moment of inertia as a function of the mass for the anisotropic interacting quark stars presented in Fig.~\ref{FigMassRadius}, where an increase in the values of $\beta_{\rm H}$ and $\beta_{\rm BL}$ implies larger moments of inertia. Additionally note that, regardless of the anisotropy model, the changes in the moment of inertia due to anisotropic pressure are negligible in the small-mass region. }
    \label{FigMomentInerMass}
\end{figure*}

Before determining the moment of inertia of the anisotropic configurations presented in Fig.~\ref{FigMassRadius}, we will analyze the radial behavior of the dragging angular velocity of a particle falling freely from infinity towards a slowly rotating star. Given a value of central density $\rho_c= 0.5 \times 10^{18}\, \rm kg/m^3$, Fig.~\ref{FigAngVelo} shows the numerical solution of Eq.~(\ref{OmegaEq}) inside and outside the star for both anisotropy models, where a positive anisotropy (with positive values of $\beta_{\rm H}$ and $\beta_{\rm BL}$) leads to increase $\omega$, while the opposite occurs for negative anisotropies. The black curves correspond to isotropic solutions, and the dragging effects on the particle are smaller as the radial distance is sufficiently large, i.e.~$\omega \rightarrow 0$ at infinity. Once we know the solution $\varpi(r)$, we can find the moment of inertia through (\ref{MomInerEq}). According to Table \ref{table1}, given a central density value, $I$ increases (decreases) as the anisotropies become more positive (negative). By repeating this procedure for a set of central densities, we can plot the moment of inertia as a function of mass, as presented in Fig.~\ref{FigMomentInerMass}. One more time, we note that the effects of anisotropy become stronger only in the high-mass region.

\begin{figure*}
    \centering
    \includegraphics[width= 8.8cm]{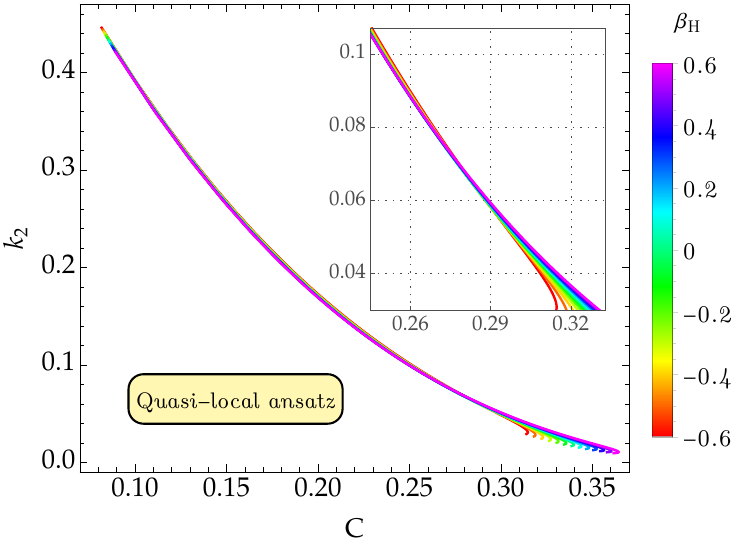}
    \includegraphics[width= 8.8cm]{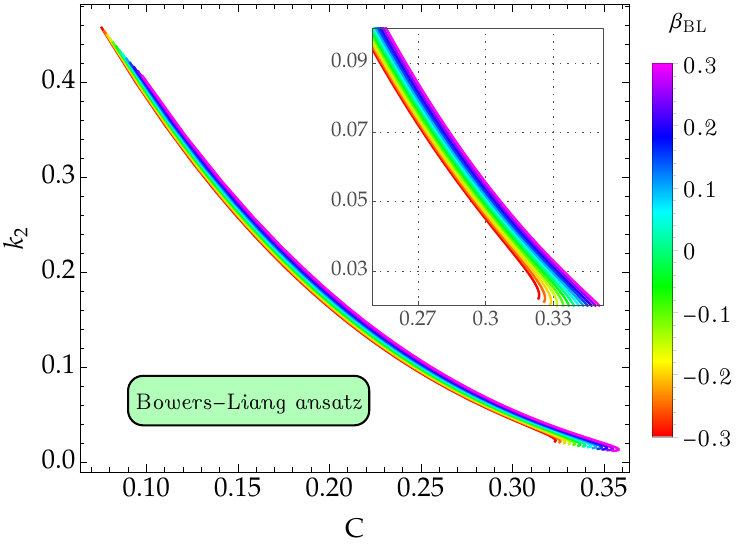}
    \caption{Tidal Love number (\ref{LoveNumEq}) versus compactness for the quasi-local ansatz (left panel) and Bowers-Liang model (right panel). For the first anisotropy model the changes in $k_2$ are irrelevant, while the anisotropic pressure in the second model generates more significant variations. In particular, positive (negative) values of the parameter $\beta_{\rm BL}$ increase (decrease) the tidal Love number of interacting QSs. }
    \label{FigLoveNumberC}
\end{figure*}

\begin{figure*}
    \centering
    \includegraphics[width= 8.8cm]{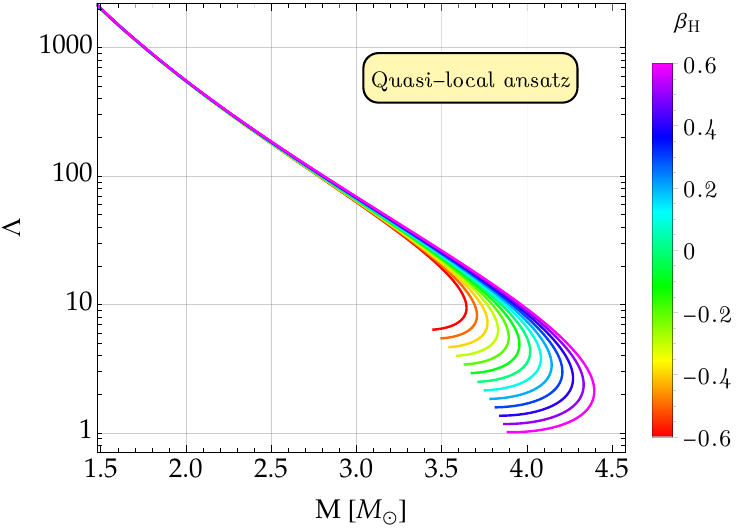}
    \includegraphics[width= 8.8cm]{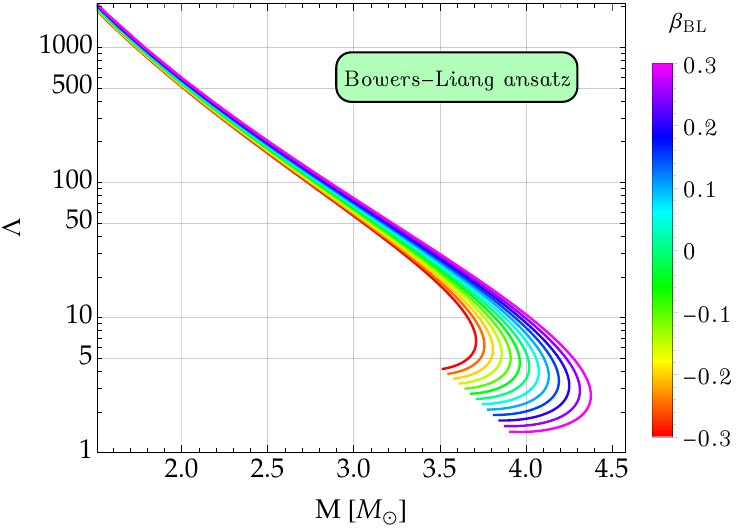}
    \caption{Dimensionless tidal deformability (\ref{TidalDeforEq}) as a function of the gravitational mass for both anisotropy profiles. One can see that the greatest impact of anisotropy on tidal deformability takes place in the high-mass branch. }
    \label{FigLambdaM}
\end{figure*}

Our next step is to investigate the dependence of the tidal Love number on the anisotropic pressure. Given an anisotropy profile and a central density value, this involves solving Eq.~(\ref{yEq}), determining $\alpha$ and therefore $k_2$ via Eq.~(\ref{LoveNumEq}). Given a value of $\rho_c$, the numerical results shown in Table \ref{table1} reveal that $k_2$ decreases with the increase of $\beta_{\rm H}$ and $\beta_{\rm BL}$. In Fig.~\ref{FigLoveNumberC} we present the tidal Love number as a function of compactness for both anisotropy models. For the quasi-local ansatz, the impact of anisotropy on $k_2$ is negligible. Specifically, in the small-compactness region, $k_2$ decreases for more positive values of $\beta_{\rm H}$, but this behavior is reversed for large compactnesses (see inset on the left plot). Nonetheless, the effect of anisotropic pressure on $k_2$ is larger for the Bowers-Liang model, and the tidal Love number always increases with increasing $\beta_{\rm BL}$. Furthermore, the dimensionless tidal deformability (\ref{TidalDeforEq}) as a function of gravitational mass is plotted in Fig.~\ref{FigLambdaM}. Once again we can observe that the most relevant effects on $\Lambda$ due to anisotropy occur in the high-mass region.

\begin{figure*}
    \centering
    \includegraphics[width= 16cm]{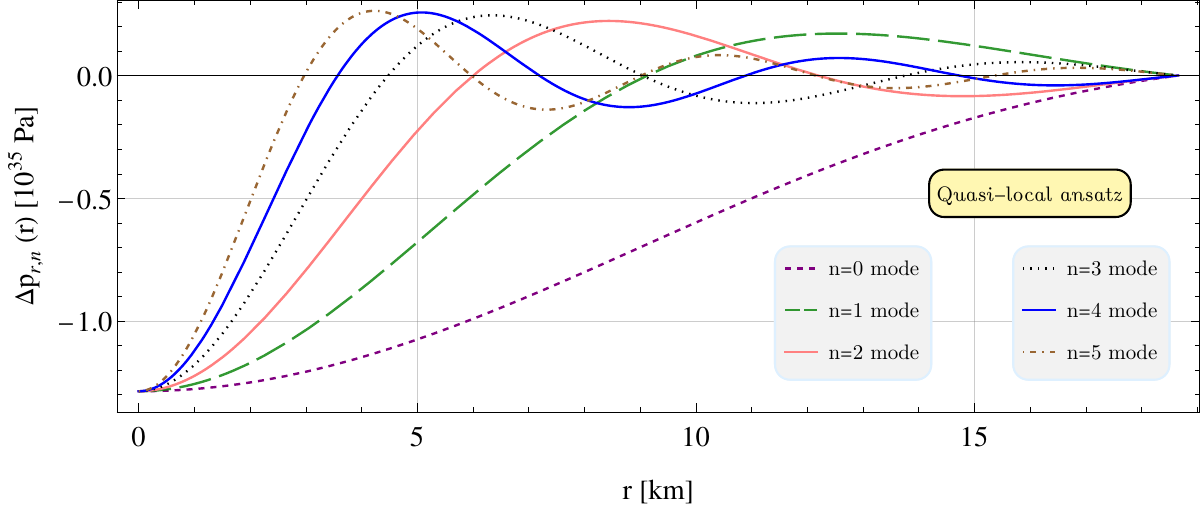}
    \includegraphics[width= 16cm]{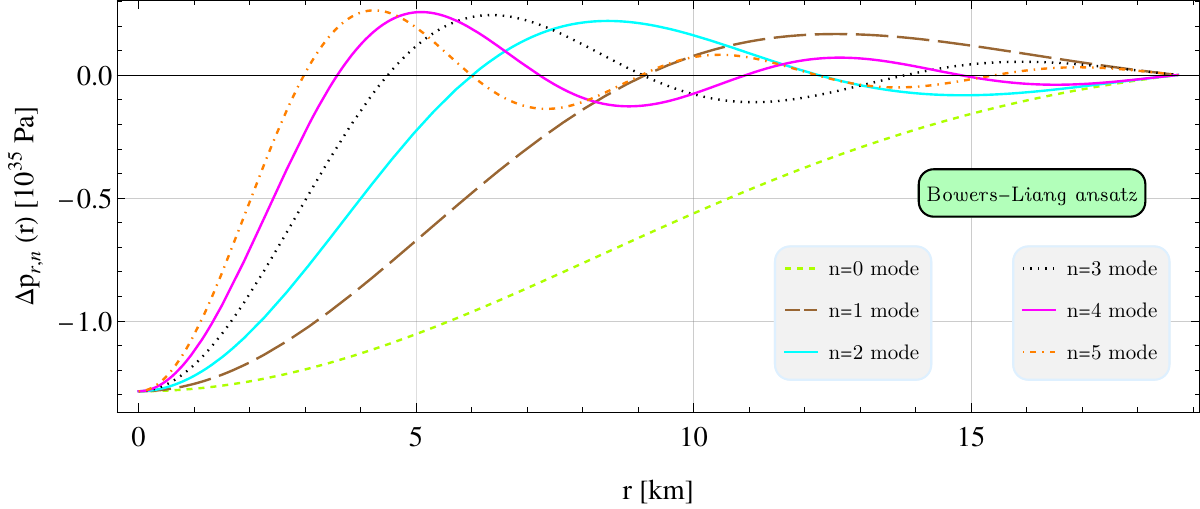}
    \caption{Radial pulsations in the interior of an interacting quark star in the presence of anisotropy for the quasi-local model with $\beta_{\rm H}= 0.4$ (upper panel) and Bowers-Liang model with $\beta_{\rm BL}= 0.2$ (lower panel). Here we have used $\rho_c= 0.5 \times 10^{18}\, \rm kg/m^3$ and EoS (\ref{eos_tot}) for $B_{\rm eff}= 60\, \rm MeV/fm^3$ and $\Bar{\lambda}= 10$. Each curve corresponds to a specific vibration mode, so that the $n$th mode contains $n$ nodes between the center and the surface of the star. }
    \label{FigModes}
\end{figure*}

\begin{figure*}
    \centering
    \includegraphics[width= 8.5cm]{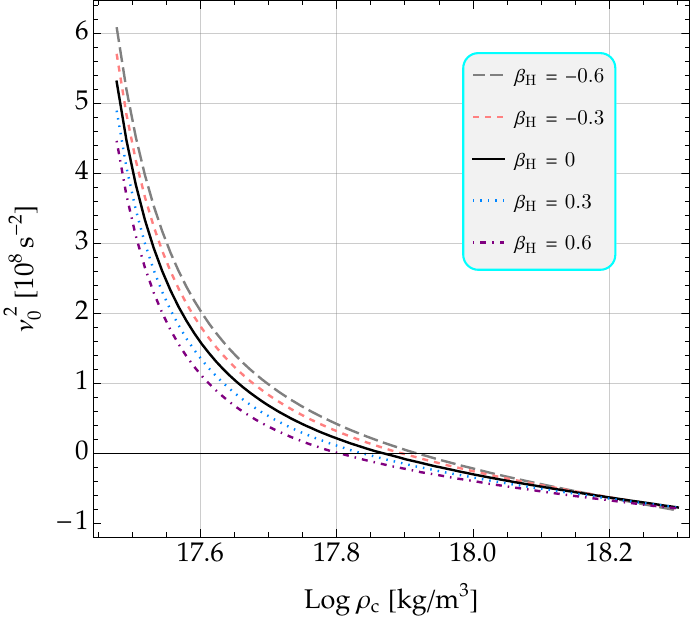}
    \includegraphics[width= 8.56cm]{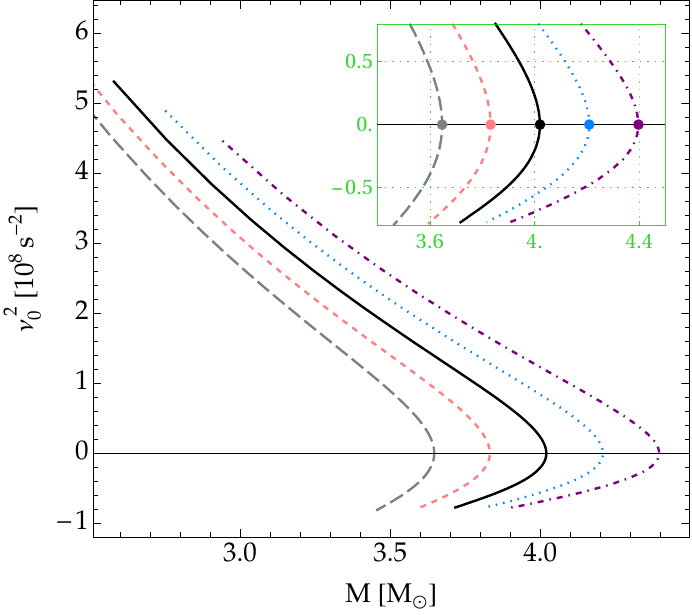}
    \caption{Squared frequency of the fundamental vibration mode as a function of the central density (left panel) and gravitational mass (right panel) for anisotropic interacting quark stars predicted by the quasi-local profile (\ref{aniso_mod1}). According to the left plot, the central density value corresponding to $\nu_0^2= 0$ is found to be an increasing value as $\beta_{\rm H}$ decreases. This means that negative anisotropies increase the radial stability of IQSs. Furthermore, according to the right plot, the maximum mass is reached exactly when $\nu_0^2=0$, regardless of the value of $\beta_{\rm H}$. Notice that filled circles in the inner box mark the maximum-mass configurations. }
    \label{FigOmega2dHorvat}
\end{figure*}

\begin{figure*}
    \centering
    \includegraphics[width= 8.50cm]{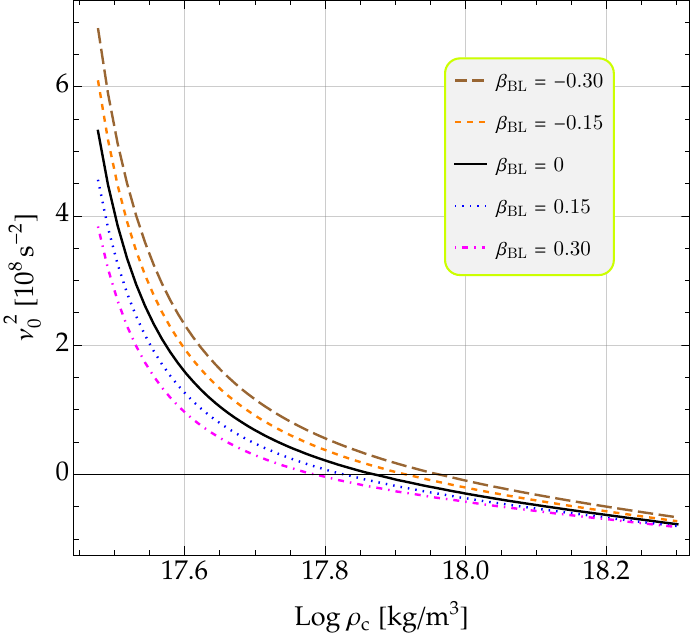}
    \includegraphics[width= 8.58cm]{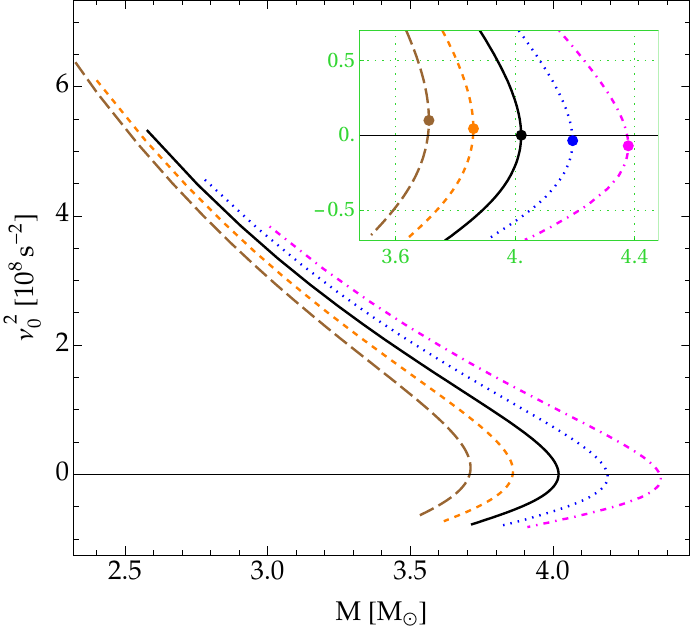}
    \caption{Squared oscillation frequencies as in Fig.~\ref{FigOmega2dHorvat}, but now for the anisotropy model proposed by Bowers and Liang (\ref{aniso_mod2}). As in the first model, $\rho_c$ corresponding to $\nu_0^2= 0$ increases as $\beta_{\rm BL}$ becomes more negative. However, according to the right plot, the maximum mass is no longer found at $\nu_0^2= 0$ for $\beta_{\rm BL} \neq 0$. The existence of stable stars beyond the maximum mass for negative anisotropies is possible. }
    \label{FigOmega2dBL}
\end{figure*}

Finally, to know whether the anisotropic configurations constructed in this work are dynamically stable, it is necessary to analyze the radial vibration modes. We begin our analysis by solving the differential equations (\ref{ROEq1}) and (\ref{ROEq2}) for a value of central density $\rho_c= 0.5 \times 10^{18}\, \rm kg/m^3$, see Fig.~\ref{FigModes}. The upper (lower) plot of this figure corresponds to the first (second) anisotropy model for $\beta_{\rm H}= 0.4$ ($\beta_{\rm BL}= 0.2$), where we have shown the eigenfunctions $\Delta p_{r,n}$ for the lowest six normal oscillation modes. For this stellar configuration, each $n$th vibration mode is associated with an eigenvalue or squared frequency $\nu_n^2$ so that we obtain a frequency spectrum of the form $\nu_0^2< \nu_1^2< \nu_2^2 < \cdots$, where $n=0$ indicates the fundamental mode and it has no nodes, $n=1$ is the first overtone (has one node), and so forth. Purely real frequencies describe dynamically stable stars (namely, when $\nu_n^2>0$), while imaginary frequencies describe unstable stars. For the sequence of equilibrium configurations presented in Fig.~\ref{FigMassRadius} with five values of $\beta_{\rm H}$, we have calculated the squared frequencies of the fundamental pulsation mode $\nu_0^2$ for the quasi-local profile (\ref {aniso_mod1}) and are shown in Fig.~\ref{FigOmega2dHorvat} as a function of the central density (left panel) and total mass (right panel). According to the right plot (see inset), the maximum mass corresponds precisely to $\nu_0^2= 0$, which means that the maximum-mass point can be used as an indicator of the transition from stability to instability of interacting quark stars with anisotropies described by the quasi-local profile. However, the right plot of Fig.~\ref{FigOmega2dBL} shows that the maximum-mass points do not coincide with $\nu_0^2=0$ for $\beta_{\rm BL} \neq 0$. Therefore, the usual stability criterion $dM/d\rho_c>0$ is no longer valid for the Bowers-Liang model. This result is similar to that obtained for anisotropic neutron stars \cite{Pretel:2020xuo}. Indeed, negative anisotropies increase the radial stability of interacting quark stars in the sense that it is possible to obtain stable stars slightly beyond the maximum mass for $\beta_{\rm BL}< 0$. Note further that, for both anisotropy models, the squared frequencies of the fundamental mode decrease with increasing central density.

\section{Theoretical prediction and observational data}\label{sec6}

\begin{figure*}
    \centering
    \includegraphics[width= 8.8cm]{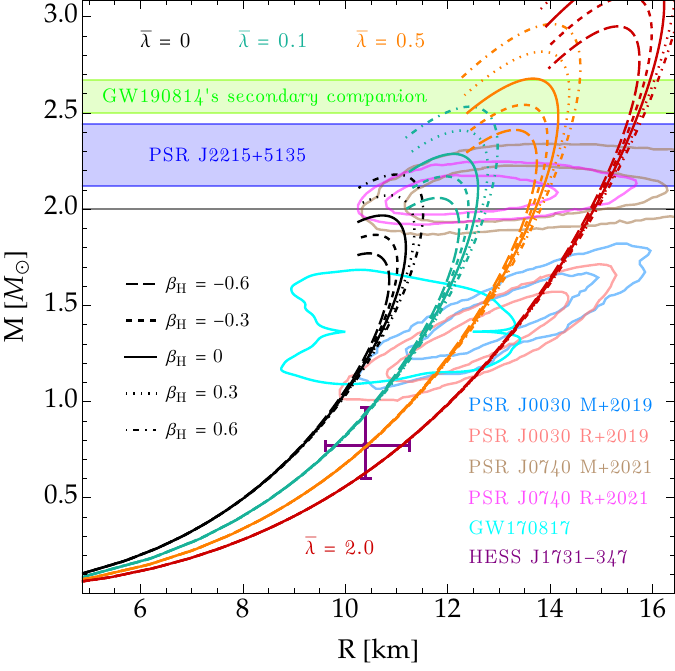}\,
    \includegraphics[width= 8.8cm]{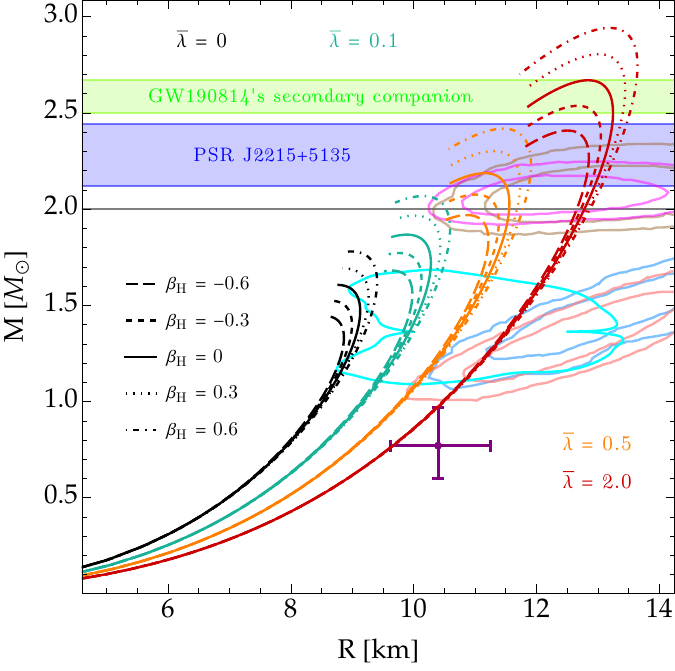}
    \caption{Comparison of mass-radius theoretical results for anisotropic interacting quark stars with observational data using the quasi-local model (\ref{aniso_mod1}) and four values of $\bar{\lambda}= \{0, 0.1, 0.5, 2.0\}$ for $B_{\rm eff}= 60\, \rm MeV/fm^3$ (left panel) and $B_{\rm eff}= 90\, \rm MeV/fm^3$ (right panel). For each value of $\bar{\lambda}$ we have considered five values of $\beta_{\rm H}$, where the isotropic case corresponds to $\beta_{\rm H}= 0$. The gray line at $2.0\, M_\odot$ represents the two massive NS pulsars J1614$-$2230 \cite{Demorest2010} and J0348$+$0432 \cite{Antoniadis2013}. The blue horizontal band indicates the observational mass measurement of the massive NS pulsar J2215$+$5135 \cite{Linares2018}, while the filled green stripe stands for the lower mass of the compact object detected by the GW 190814 event \cite{Abbott2020AJL}. The purple dot with its respective error bars represents the supernova remnant HESS J1731$-$347 \cite{Doroshenko2022} and the cyan region is the mass-radius constraint from the GW170817 event \cite{Abbott2018PRL}. Moreover, NICER constraints obtained from the pulsars PSR J0030$+$0451 \cite{Miller:2019cac, Riley2019} and PSR J0740$+$6620 \cite{Miller2021, Riley2021} are indicated by different color contours. Note that, as a consequence of increasing $B_{\rm eff}$, the radius and mass of the maximum-mass configuration decrease. }
    \label{FigObsDataMR}
\end{figure*}

Astrophysical measurements, based on observations of NSs in the X-ray band and on gravitational wave observations, play a crucial role in adopting or discarding an EoS. In that perspective, it is essential to compare our theoretical results with the recent observational measurements/constraints. In Fig.~\ref{FigObsDataMR}, we compare our mass-radius results with the NICER measurements obtained from the pulsars PSR J0030$+$0451 \cite{Miller:2019cac, Riley2019} and PSR J0740$+$6620 \cite{Miller2021, Riley2021}. We have included the constraint from the GW170817 event \cite{Abbott2018PRL}, as well as the mass of the secondary component of the gravitational-wave signal GW190814 \cite{Abbott2020AJL}. We also incorporate the central compact object within the supernova remnant HESS J1731$-$347, whose mass and radius estimates have been reported in Ref.~\cite{Doroshenko2022}. All these measurements are being represented by different colored bands or contours in both plots of Fig.~\ref{FigObsDataMR}.

We begin our comparative analysis with the noninteracting case (i.e., when $\bar{\lambda} = 0$) for $B_{\rm eff}= 60\, \rm MeV/fm^3$, see black curves in the left panel of Fig.~\ref{FigObsDataMR}. As one can observe, isotropic solutions ($\beta_{\rm H}= 0$) do not provide astrophysical masses above $2\, M_{\odot}$, which does not favor the description of highly massive pulsars. However, for positive anisotropies it is possible to exceed two solar masses, see the exemplary cases when $\beta_{\rm H}= 0.3$ and $0.6$. Remarkably, when strong interaction effects take place (i.e., when $\bar{\lambda}> 0$), we obtain results that satisfy most observational $M-R$ measurements. In particular, for $\bar{\lambda} \in [0.1, 2.0]$, our outcomes suggest that the supernova remnant HESS J1731$-$347 can be described as an anisotropic IQS, regardless of the value of $\beta_{\rm H}$. This is allowed because the effect of anisotropy is irrelevant for small masses. Nonetheless, the anisotropic pressure plays an important role in the high-mass region by satisfying the constraints obtained from highly massive Pulsars observations, see specifically the green and orange curves. We further note that $\bar{\lambda}= 0.5$ and $\beta_{\rm H} \in [-0.4, 0.6]$ favor a consistent description for the secondary component resulting from the GW190814 event. Besides, in the right plot of Fig.~\ref{FigObsDataMR}, we display the $M-R$ outcomes corresponding to $B_{\rm eff}= 90\, \rm MeV/fm^3$, where we can perceive that an increase in the effective bag constant leads to a decrease in the maximum mass, but it is still possible to obtain results compatible with the observational data, see for example the red curves for $\bar{\lambda}= 2.0$.

\begin{figure}
    \centering
    \includegraphics[width= 8.5cm]{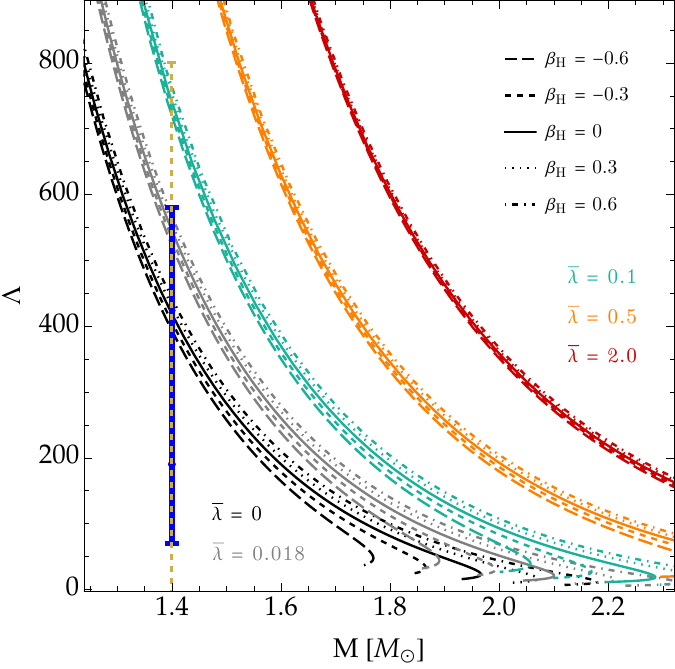}
    \caption{Dimensionless tidal deformability versus total mass relation for the stellar configurations presented in the left panel of Fig.~\ref{FigObsDataMR}, where we have additionally included the results for $\bar{\lambda}= 0.018$ by gray curves. The blue vertical line denotes $\Lambda_{1.4M_\odot}= 190_{-120}^{+390}$ reported by LIGO-Virgo Collaboration by analyzing the tidal effects of the GW170817 event \cite{Abbott2018PRL}. The dark yellow dashed line stands for the EoS-independent constraint $\Lambda_{1.4M_\odot} \leq 800$ from Ref.~\cite{Abbott2017PRL}. }
    \label{FigObsDataLM}
\end{figure}

\begin{figure}
    \centering
    \includegraphics[width= 8.5cm]{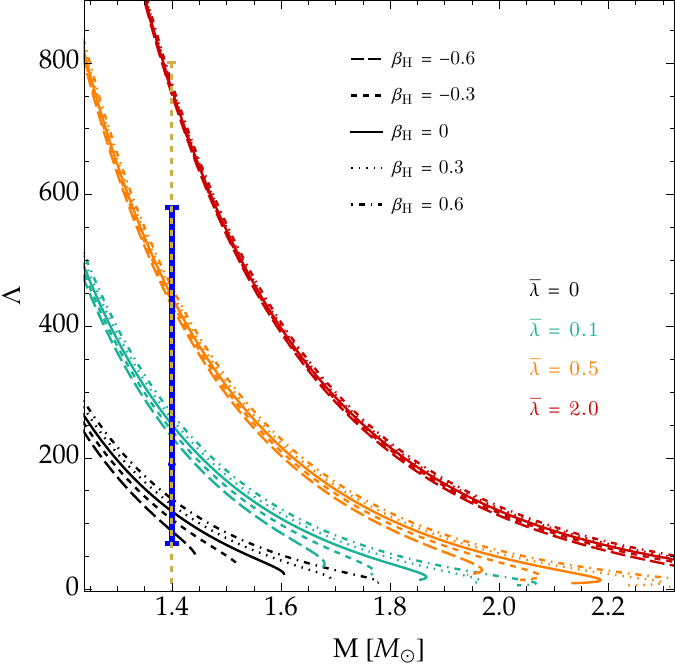}
    \caption{Tidal deformability properties for the equilibrium configurations shown in the right panel of Fig.~\ref{FigObsDataMR}, namely, when $B_{\rm eff}= 90\, \rm MeV/fm^3$. Here, all our results are consistent with the EoS-independent constraint $\Lambda_{1.4M_\odot} \leq 800$ \cite{Abbott2017PRL}. }
    \label{FigObsDataLM90}
\end{figure}

For the anisotropic IQSs presented in the left plot of Fig.~\ref{FigObsDataMR}, the dimensionless tidal deformability as a function of the total mass is shown in Fig.~\ref{FigObsDataLM}, where we have included the estimate reported by LIGO-Virgo Collaboration for the tidal effects of the GW170817 event \cite{Abbott2018PRL}, namely $\Lambda_{1.4M_\odot}= 190_{-120}^{+390}$. It can be seen that, for $\bar{\lambda}\lesssim 0.018$, any value of the anisotropy parameter $\beta_{\rm H}$ satisfies such an estimate. Nevertheless, we must point out that, this constraint is based on the assumption that the coalescing bodies were NSs. Actually, for quark stars in the present study, we must use the EoS-independent constraint $\Lambda_{1.4M_\odot} \leq 800$ (see the dark yellow dashed vertical line) from the older LIGO/Virgo paper \cite{Abbott2017PRL}, which is satisfied for $\bar{\lambda}\lesssim 0.1$ if we adopt $B_{\rm eff}= 60\, \rm MeV/fm^3$. As one can appreciate, this last restriction can further open the free parameter space for IQSs. Noticeably, when we assume $B_{\rm eff}= 90\, \rm MeV/fm^3$, all our findings are consistent with the tidal deformability constraints for $\bar{\lambda} \in [0, 2.0]$, as shown in Fig.~\ref{FigObsDataLM90}. Thus, quark stars made of interacting quark matter with large effective bag constant and proper $\bar{\lambda}$ choices can meet the tidal deformability constraint of the GW170817 event while satisfying the maximum mass restrictions, see also right panel of Fig.~\ref{FigObsDataMR}.

\section{Conclusions}\label{sec7}

Within the Standard Model of particle physics, Quantum chromodynamics (QCD) is the accepted theory of strong interaction force between quarks mediated by gluons. At the same time, perturbative expansions in QCD and color superconductivity allow us to describe interacting quark matter that includes the interquark effects induced by strong interaction. This may reveal new physical phenomena under extreme conditions such as inside compact stars. In this work, we have investigated the role of strong interaction on anisotropic QSs composed of interacting quark matter. Our study involves spherically symmetric anisotropic QSs adopting two physically well-motivated anisotropy profiles. To achieve this end, we have numerically solved the stellar structure equations in Einstein gravity using suitable boundary conditions and we discussed various global properties such as mass-radius relation, surface redshift, moment of inertia, tidal properties and oscillation spectrum.

The local anisotropy has been included with the help of two free parameter $\beta_{\rm H}$ and $\beta_{\rm BL}$ for the adopted models. By choosing particular values of these parameters, we have obtained the $M-R$ relations for anisotropic interacting quark stars, as illustrated in Fig.~\ref{FigMassRadius}. We noticed that the mass and radius simultaneously increase for positive increasing values of anisotropy factor, and this holds for both models, see Table \ref{table1}. According to the $M(\rho_c)$ method in the $M-\rho_c$ diagram in Fig.~\ref{FigMassCentDen}, where the maximum-mass point indicates that stability ceases, the critical central density (corresponding to the maximum mass) decreases with the increase of the parameters $\beta_{\rm H}$ and $\beta_{\rm BL}$. In other words, a positive (negative) anisotropy decreases (increases) the stability of IQSs. Depending on the obtained values of mass and radius, we also determined the gravitational redshift $z_{\rm sur}$, which is strongly affected by the presence of anisotropy in the high-mass region.

Our next step was to analyze the behavior of the frame-dragging angular velocity and moment of inertia in the slowly rotating approximation. In particular, given a central density value, we have shown that a positive anisotropy leads to significantly increasing the angular velocity mainly in the innermost regions of the IQSs, while the opposite occurs for negative anisotropies. The anisotropic pressure consequently increases the moment of inertia of the stars analyzed as $\beta_{\rm H}$ and $\beta_{\rm BL}$ increase from their negative values, showing its greatest impact on the most massive stars.

In our analysis of tidal properties we found that the tidal Love number $k_2$ is weakly altered by anisotropy for the model proposed by Horvat \textit{et al.}, while the changes are more pronounced for the Bowers-Liang model. Specifically, positive (negative) values of the parameter $\beta_{\rm BL}$ increase (decrease) the tidal Love number of IQSs. Moreover, the main effect of anisotropy on the dimensionless tidal deformability $\Lambda$ takes place in the high-mass branch.

The already mentioned $M(\rho_c)$ method provides a simple and necessary condition for stellar stability, however, a sufficient condition is to determine the frequencies of normal pulsation modes when a star is perturbed radially and adiabatically. So we have also performed a rigorous examination of the anisotropic stellar configurations constructed here following a radial oscillations approach for each anisotropy profile. Noticeably, for both anisotropy models, the central density value corresponding to zero squared frequency of the fundamental vibration mode (namely, $\nu_0^2= 0$) was found to be an increasing value as $\beta_{\rm H}$ and $\beta_{\rm BL}$ decrease. In other words, negative anisotropies increase the radial stability of IQSs, as predicted by the $M(\rho_c)$ method. For the quasi-local model, the $M(\rho_c)$ method is compatible with the calculation of radial oscillation frequencies, so the maximum mass can be used as an indicator of the onset of instability for any value of $\beta_{\rm H}$. Notwithstanding, for the Bowers-Liang profile, the central density corresponding to the maximum-mass point does not coincide with the central density where the squared oscillation frequency vanishes, indicating that the existence of stable anisotropic IQSs is possible beyond the maximum mass for negative anisotropies. Meanwhile, stars stop being stable before reaching the maximum mass for positive values of $\beta_{\rm BL}$. This is an important outcome that substantially differentiates one anisotropy model from the other.

Finally, we have carried out a comparative analysis between our theoretical results and observational data. It is important to highlight that, when strong interaction effects take place (i.e., for $\bar{\lambda}> 0$), we obtained results that satisfy most observational mass-radius measurements. Specifically, when $\bar{\lambda} \in [0.1, 2.0]$, our calculations suggest that the central compact object within the supernova remnant HESS J1731$-$347 can be described as an anisotropic interacting quark star, regardless of the value of $\beta_{\rm H}$. Furthermore, according to the outcomes shown in Figs.~\ref{FigMassRadius}-\ref{FigLambdaM}, we observe that the anisotropy profiles yield identical quark-star characteristics, at least for the range of values of $\beta_{\rm H}$ and $\beta _{\rm BL}$ considered in this study. Thus, with the exception of the radial stability analysis, we must remark that the effect of anisotropy generated by both models on IQSs is similar. However, as shown in Refs.~\cite{Pretel:2020xuo, Rahmansyah:2021gzt} from the mass-radius diagrams of anisotropic NSs, there is a notable difference between the findings of the two anisotropy profiles not only for high masses but also in the low-mass region. Here, our work has shown that this difference is more subtle for interacting quark matter, where the $M-R$ relation is almost unalterable due to anisotropy in the smaller mass branch. Remarkably, our findings are comparable and consistent with the tidal deformability constraint from the GW170817 event \cite{Abbott2017PRL} while satisfying the different observational mass-radius measurements.

\begin{acknowledgments}
We would like to thank Chen Zhang for his helpful and constructive comments on the interacting quark matter equation of state. JMZP acknowledges financial support from the PCI program of the Brazilian agency ``Conselho Nacional de Desenvolvimento Cient{\'i}fico e Tecnol{\'o}gico''--CNPq. T.~Tangphati is supported by School of Science, Walailak University, Thailand. A. Pradhan thanks to IUCCA, Pune, India for providing facilities under associateship programmes.
\end{acknowledgments}\

%%%%%%%%%%%%%%%%%%%%%%%%%%%%%%%%%%%%%%%%%%%%%%%%%%%%%%%%%%%%%%%%%%


\begin{thebibliography}{90}

\bibitem{Will:2014kxa}
C.~M.~Will,
%``The Confrontation between General Relativity and Experiment,''
\href{https://doi.org/10.12942/lrr-2014-4}{Living Rev. Rel. \textbf{17}, 4 (2014)}.

\bibitem{Ozel:2016oaf}
F.~\"Ozel and P.~Freire,
%``Masses, Radii, and the Equation of State of Neutron Stars,''
\href{https://doi.org/10.1146/annurev-astro-081915-023322}{Ann. Rev. Astron. Astrophys. \textbf{54}, 401 (2016)}.

\bibitem{Steiner:2017vmg}
A.~W.~Steiner \textit{et al.},
%``Constraining the Mass and Radius of Neutron Stars in Globular Clusters,''
\href{https://doi.org/10.1093/mnras/sty215}{MNRAS \textbf{476}, 421 (2018)}.

\bibitem{Miller:2019cac}
M.~C.~Miller \textit{et al.}
%``PSR J0030+0451 Mass and Radius from $NICER$ Data and Implications for the Properties of Neutron Star Matter,''
\href{https://doi.org/10.3847/2041-8213/ab50c5}{Astrophys. J. Lett. \textbf{887}, L24 (2019)}.




\bibitem{Itoh}
N. Itoh,
%``Hydrostatic Equilibrium of Hypothetical Quark Stars,'' 
\href{https://doi.org/10.1143/PTP.44.291}{Progr. Theor. Phys. \textbf{44}, 291 (1970)}.

\bibitem{Witten:1984rs}
E.~Witten,
%``Cosmic Separation of Phases,''
\href{https://doi.org/10.1103/PhysRevD.30.272}{Phys. Rev. D \textbf{30}, 272 (1984)}.

\bibitem{Bodmer:1971we}
A.~R.~Bodmer,
%``Collapsed nuclei,''
\href{https://doi.org/10.1103/PhysRevD.4.1601}{Phys. Rev. D \textbf{4}, 1601 (1971)}.

\bibitem{Farhi:1984qu}
E.~Farhi and R.~L.~Jaffe,
%``Strange Matter,''
\href{https://doi.org/10.1103/PhysRevD.30.2379}{Phys. Rev. D \textbf{30}, 2379 (1984)}.

\bibitem{Nicotra:2006eg}
O.~E.~Nicotra, M.~Baldo, G.~F.~Burgio and H.~J.~Schulze,
%``Hybrid protoneutron stars with the MIT bag model,''
\href{https://doi.org/10.1103/PhysRevD.74.123001}{Phys. Rev. D \textbf{74}, 123001 (2006)}.

\bibitem{Arbanil:2016wud}
J.~D.~V.~Arba\~nil and M.~Malheiro,
%``Radial stability of anisotropic strange quark stars,''
\href{https://doi.org/10.1088/1475-7516/2016/11/012}{JCAP \textbf{11}, 012 (2016)}.

\bibitem{Joshi:2020lwn}
S.~Joshi, S.~Sau and S.~Sanyal,
%``Quark cores in extensions of the MIT Bag model,''
\href{https://doi.org/10.1016/j.jheap.2021.03.001}{JHEAp \textbf{30}, 16 (2021)}.

\bibitem{Lopes2021PS}
L.~L.~Lopes, C.~Biesdorf and D.~P.~Menezes,
%``MModified MIT bag Models—part I: Thermodynamic consistency, stability windows and symmetry group,''
\href{https://doi.org/10.1088/1402-4896/abef34}{Phys. Scripta \textbf{96}, 065303 (2021)}.

\bibitem{Lopes:2020dvs}
L.~L.~Lopes, C.~Biesdorf, K.~D.~Marquez and D.~P.~Menezes,
%``Modified MIT Bag Models -- part II: QCD phase diagram and hot quark stars,''
\href{https://doi.org/10.1088/1402-4896/abef35}{Phys. Scripta \textbf{96}, 065302 (2021)}.

\bibitem{Arbanil2023}
J.~D.~V.~Arba{\~n}il, C.~V.~Flores, C.~H.~Lenzi and J.~M.~Z.~Pretel,
%``Fluid pulsation modes and tidal deformability of anisotropic strange stars in light of the GW170817 event,''
\href{https://doi.org/10.1103/PhysRevD.107.124016}{Phys. Rev. D \textbf{107}, 124016 (2023)}.



\bibitem{Astashenok2015}
A.~V.~Astashenok, S.~Capozziello and S.~D.~Odintsov,
%``Nonperturbative models of quark stars in $f(R)$ gravity,''
\href{https://doi.org/10.1016/j.physletb.2015.01.030}{Phys. Lett. B \textbf{742}, 160 (2015)}.

\bibitem{Salako:2021xkj}
I.~G.~Salako, D.~R.~Boko, G.~F.~Pomalegni and M.~Z.~Arouko,
%``Study on anisotropic strange stars in Rastall gravity,''
\href{https://doi.org/10.1134/s0040577921090105}{Theor. Math. Phys. \textbf{208}, 1299 (2021)}.

\bibitem{Pretel:2022plg}
J.~M.~Z.~Pretel and S.~B.~Duarte,
%``Anisotropic quark stars in $f(R) = R^{1+ϵ}$ gravity,''
\href{https://doi.org/10.1088/1361-6382/ac7a88}{Class. Quant. Grav. \textbf{39}, 155003 (2022)}.

\bibitem{PretelTA2023}
J.~M.~Z. Pretel, T.~Tangphati and A.~Banerjee,
%``Relativistic structure of charged quark stars in energy-momentum squared gravity,''
\href{https://doi.org/10.1016/j.aop.2023.169440}{Ann. Phys. \textbf{458}, 169440 (2023)}.




\bibitem{Roupas2021}
Z. Roupas,
%``Secondary component of gravitational-wave signal GW190814 as an anisotropic neutron star,''
\href{https://doi.org/10.1007/s10509-021-03919-5}{Astrophys. Space Sci. \textbf{366}, 9 (2021)}.

\bibitem{Bowers:1974tgi}
R.~L.~Bowers and E.~P.~T.~Liang,
%``Anisotropic Spheres in General Relativity,''
\href{https://doi.org/10.1086/152760}{Astrophys. J. \textbf{188}, 657 (1974)}.

\bibitem{Herrera:1994qua}
L.~Herrera and N.~O.~Santos,
%``Jeans mass for anisotropic matter,''
\href{https://doi.org/10.1086/175075}{Astrophys. J. \textbf{438}, 308 (1994)}.

\bibitem{Mak:2001eb}
M.~K.~Mak and T.~Harko,
%``Anisotropic stars in general relativity,''
\href{https://doi.org/10.1098/rspa.2002.1014}{Proc. Roy. Soc. Lond. A \textbf{459}, 393 (2003)}.

\bibitem{Lake:2009cd}
K.~Lake,
%``Generating static spherically symmetric anisotropic solutions of Einstein's equations from isotropic Newtonian solutions,''
\href{https://doi.org/10.1103/PhysRevD.80.064039}{Phys. Rev. D \textbf{80}, 064039 (2009)}.

\bibitem{Ivanov:2017kyr}
B.~V.~Ivanov,
%``Analytical study of anisotropic compact star models,''
\href{https://doi.org/10.1140/epjc/s10052-017-5322-7}{Eur. Phys. J. C \textbf{77}, 738 (2017)}.

\bibitem{Stelea:2018cgm}
C.~Stelea, M.~A.~Dariescu and C.~Dariescu,
%``Magnetized anisotropic stars,''
\href{https://doi.org/10.1103/PhysRevD.97.104059}{Phys. Rev. D \textbf{97}, 104059 (2018)}.

\bibitem{Biswas2019}
B. Biswas and S. Bose,
%``Tidal deformability of an anisotropic compact star: Implications of GW170817,''
\href{https://doi.org/10.1103/PhysRevD.99.104002}{Phys. Rev. D \textbf{99}, 104002 (2019)}.

\bibitem{DasCPC2023}
B. Das, K. B. Goswami, A. Saha and P. K. Chattopadhyay,
%``Anisotropic strange quark star in Finch-Skea geometry and its maximum mass for non-zero strange quark mass (ms ≠ 0),''
\href{https://doi.org/10.1088/1674-1137/acb90f}{Chinese Phys. C \textbf{47}, 055101 (2023)}.

\bibitem{Baskey2023}
L. Baskey \textit{et al.},
%``Anisotropic compact stellar solution in general relativity,''
\href{https://doi.org/10.1140/epjc/s10052-023-11351-y}{Eur. Phys. J. C \textbf{83}, 307 (2023)}.

\bibitem{Parida2023}
B. K. Parida, S. Das and M. Govender,
%``Toy models of compact anisotropic stars and their Love numbers,''
\href{https://doi.org/10.1142/S0218271823500384}{Int. J. Mod. Phys. D \textbf{32}, 2350038 (2023)}.









\bibitem{Horvat:2010xf}
D.~Horvat, S.~Ilijic and A.~Marunovic,
%``Radial pulsations and stability of anisotropic stars with quasi-local equation of state,''
\href{https://doi.org/10.1088/0264-9381/28/2/025009}{Class. Quant. Grav. \textbf{28}, 025009 (2011)}.



\bibitem{Pretel:2020xuo}
J.~M.~Z.~Pretel,
%``Equilibrium, radial stability and non-adiabatic gravitational collapse of anisotropic neutron stars,''
\href{https://doi.org/10.1140/epjc/s10052-020-8301-3}{Eur. Phys. J. C \textbf{80}, 726 (2020)}.

\bibitem{Curi:2022nnt}
E.~J.~A.~Curi, L.~B.~Castro, C.~V.~Flores and C.~H.~Lenzi,
%``Non-radial oscillations and global stellar properties of anisotropic compact stars using realistic equations of state,''
\href{https://doi.org/10.1140/epjc/s10052-022-10498-4}{Eur. Phys. J. C \textbf{82}, 527 (2022)}.

\bibitem{Das:2023pfq}
H.~C.~Das, J.~A.~Pattnaik and S.~K.~Patra,
%``Anisotropy effects on the neutron star properties,''
DAE Symp. Nucl. Phys. \textbf{66}, 764 (2023).

\bibitem{Das:2023jtj}
H.~C.~Das, J.~A.~Pattnaik and S.~K.~Patra,
%``Constraining the surface curvature of an anisotropic neutron star,''
\href{https://doi.org/10.1103/PhysRevD.107.083007}{Phys. Rev. D \textbf{107}, 083007 (2023)}.

\bibitem{Pretel2023}
J.~M.~Z.~Pretel,
%``Radial pulsations, moment of inertia and tidal deformability of dark energy stars,''
\href{https://doi.org/10.1140/epjc/s10052-023-11198-3}{Eur. Phys. J. C \textbf{83}, 26 (2023)}.

\bibitem{Pattersons:2021lci}
M.~L.~Pattersons and A.~Sulaksono,
%``Mass correction and deformation of slowly rotating anisotropic neutron stars based on Hartle-Thorne formalism,''
\href{https://doi.org/10.1140/epjc/s10052-021-09481-2}{Eur. Phys. J. C \textbf{81}, 698 (2021)}.


\bibitem{Silva:2014fca}
H.~O.~Silva, C.~F.~B.~Macedo, E.~Berti and L.~C.~B.~Crispino,
%``Slowly rotating anisotropic neutron stars in general relativity and scalar-tensor theory,''
\href{https://doi.org/10.1088/0264-9381/32/14/145008}{Class. Quant. Grav. \textbf{32}, 145008 (2015)}.

\bibitem{Pretel:2022qng}
J.~M.~Z.~Pretel,
%``Moment of inertia of slowly rotating anisotropic neutron stars in f(R,T) gravity,''
\href{https://doi.org/10.1142/S0217732322501887}{Mod. Phys. Lett. A \textbf{37}, 2250188 (2022)}.









%%%%%%%%%%%%%%%%%%%%%%%%%%%%%%%%%%%%%%%%%%%%



\bibitem{Zhang:2020jmb}
C.~Zhang and R.~B.~Mann,
%``Unified Interacting Quark Matter and its Astrophysical Implications,''
\href{https://doi.org/10.1103/PhysRevD.103.063018}{Phys. Rev. D \textbf{103}, 063018 (2021)}.

\bibitem{Zhang:2021fla}
C.~Zhang,
%``Gravitational wave echoes from interacting quark stars,''
\href{https://doi.org/10.1103/PhysRevD.104.083032}{Phys. Rev. D \textbf{104}, 083032 (2021)}.



\bibitem{Hartle1967}
J.~B.~Hartle,
%``Slowly Rotating Relativistic Stars. I. Equations of Structure,''
\href{https://ui.adsabs.harvard.edu/abs/1967ApJ...150.1005H}{Astrophys. J. \textbf{150}, 1005 (1967)}.

\bibitem{Chatziioannou2020}
K.~Chatziioannou,
%``Neutron-star tidal deformability and equation-of-state constraints,''
\href{https://doi.org/10.1007/s10714-020-02754-3}{Gen. Relativ. Gravit. \textbf{52}, 109 (2020)}.

\bibitem{Kanakis2020}
A.~Kanakis-Pegios, P.~S.~Koliogiannis and Ch.~C.~Moustakidis,
%``Speed of sound constraints from tidal deformability of neutron stars,''
\href{https://doi.org/10.1103/PhysRevC.102.055801}{Phys. Rev. C \textbf{102}, 055801 (2020)}.

\bibitem{Yang2023}
S.~Yang, C.~Pi, X.~Zheng and F.~Weber,
%``Confronting Strange Stars with Compact-Star Observations and New Physics,''
\href{https://doi.org/10.3390/universe9050202}{Universe \textbf{9}, 202 (2023)}.

\bibitem{Postnikov2021}
S.~Postnikov, M.~Prakash and J.~M.~Lattimer,
%``Tidal Love numbers of neutron and self-bound quark stars,''
\href{https://doi.org/10.1103/PhysRevD.82.024016}{Phys. Rev. D \textbf{82}, 024016 (2010)}.

\bibitem{Chandrasekhar1964}
S.~Chandrasekhar,
%``The Dynamical Instability of Gaseous Masses Approaching the Schwarzschild Limit in General Relativity,''
\href{https://doi.org/10.1086/147938}{Astrophys. J. \textbf{140}, 417 (1964)}.



\bibitem{Folomeev:2018ioy}
V.~Folomeev,
%``Anisotropic neutron stars in $R^2$ gravity,''
\href{https://doi.org/10.1103/PhysRevD.97.124009}{Phys. Rev. D \textbf{97}, 124009 (2018)}.

\bibitem{Doneva:2012rd}
D.~D.~Doneva and S.~S.~Yazadjiev,
%``Gravitational wave spectrum of anisotropic neutron stars in Cowling approximation,''
\href{https://doi.org/10.1103/PhysRevD.85.124023}{Phys. Rev. D \textbf{85}, 124023 (2012)}.

\bibitem{Yagi:2015hda}
K.~Yagi and N.~Yunes,
%``I-Love-Q anisotropically: Universal relations for compact stars with scalar pressure anisotropy,''
\href{https://doi.org/10.1103/PhysRevD.91.123008}{Phys. Rev. D \textbf{91}, 123008 (2015)}.

\bibitem{Rahmansyah:2020gar}
A.~Rahmansyah, A.~Sulaksono, A.~B.~Wahidin and A.~M.~Setiawan,
%``Anisotropic neutron stars with hyperons: implication of the recent nuclear matter data and observations of neutron stars,''
\href{https://doi.org/10.1140/epjc/s10052-020-8361-4}{Eur. Phys. J. C \textbf{80}, 769 (2020)}.

\bibitem{Rahmansyah:2021gzt}
A.~Rahmansyah and A.~Sulaksono,
%``Recent multimessenger constraints and the anisotropic neutron star,''
\href{https://doi.org/10.1103/PhysRevC.104.065805}{Phys. Rev. C \textbf{104}, 065805 (2021)}.







%%%%%%%%%%%%%%%%%%%%%%%%%%%%%%%%%%%%%%%%%%%%%%%%%%%%%%%%%%%%%




\bibitem{Mak2003}
M. K. Mak and T. Harko,
%``Anisotropic stars in general relativity,''
\href{https://doi.org/10.1098/rspa.2002.1014}{Proc. R. Soc. Lond. A \textbf{459}, 393 (2003)}.

\bibitem{Demorest2010}
P. Demorest \textit{et al.},
%``A two-solar-mass neutron star measured using Shapiro delay,''
\href{https://doi.org/10.1038/nature09466}{Nature \textbf{467}, 1081 (2010)}.

\bibitem{Antoniadis2013}
J. Antoniadis \textit{et al.},
%``A massive pulsar in a compact relativistic binary,''
\href{https://doi.org/10.1126/science.1233232}{Science \textbf{340}, 6131 (2013)}.

\bibitem{Linares2018}
M. Linares, T. Shahbaz and J. Casares,
%``Peering into the dark side: magnesium lines establish a massive neutron star in PSR J2215+5135,''
\href{https://doi.org/10.3847/1538-4357/aabde6}{Astrophys. J. \textbf{859}, 54 (2018)}.

\bibitem{Abbott2020AJL}
R. Abbott \textit{et al.},
%``GW190814: gravitational waves from the coalescence of a 23 solar mass black hole with a 2.6 solar mass compact object,''
\href{https://doi.org/10.3847/2041-8213/ab960f}{Astrophys. J. Lett. \textbf{896}, L44 (2020)}.

\bibitem{Doroshenko2022}
V. Doroshenko, V. Suleimanov, G. Phlhofer and A. Santangelo,
%``A strangely light neutron star within a supernova remnant,''
\href{https://doi.org/10.1038/s41550-022-01800-1}{Nat. Astron. \textbf{6}, 1444 (2022)}.

\bibitem{Abbott2018PRL}
B. P. Abbott \textit{et al.},
%``GW170817: Measurements of Neutron Star Radii and Equation of State,''
\href{https://doi.org/10.1103/PhysRevLett.121.161101}{Phys. Rev. Lett. \textbf{121}, 161101 (2018)}.

\bibitem{Riley2019}
T. E. Riley \textit{et al.},
%``A NICER View of PSR J0030+0451: Millisecond Pulsar Parameter Estimation,''
\href{https://doi.org/10.3847/2041-8213/ab481c}{Astrophys. J. Lett. \textbf{887}, L21 (2019)}.

\bibitem{Miller2021}
M. C. Miller \textit{et al.},
%``The Radius of PSR J0740+6620 from NICER and XMM-Newton Data,''
\href{https://doi.org/10.3847/2041-8213/ac089b}{Astrophys. J. Lett. \textbf{918}, L28 (2021)}.

\bibitem{Riley2021}
T. E. Riley \textit{et al.},
%``A NICER View of the Massive Pulsar PSR J0740+6620 Informed by Radio Timing and XMM-Newton Spectroscopy,''
\href{https://doi.org/10.3847/2041-8213/ac0a81}{Astrophys. J. Lett. \textbf{918}, L27 (2021)}.

\bibitem{Abbott2017PRL}
B. P. Abbott \textit{et al.},
%``GW170817: Observation of Gravitational Waves from a Binary Neutron Star Inspiral,''
\href{https://doi.org/10.1103/PhysRevLett.119.161101}{Phys. Rev. Lett. \textbf{119}, 161101 (2017)}.








\end{thebibliography}
\end{document}